\documentclass[a4paper,usenatbib]{mnras}

\usepackage{graphicx}
\usepackage{amssymb}

\title[Time-resolved WISE Coadds]{Time-resolved WISE/NEOWISE Coadds}

\author[Meisner et al.]{
A.~M. Meisner,$^{1,2}$\thanks{ameisner@lbl.gov}
D. Lang$^{3,4}$
and D.~J. Schlegel$^{2}$
\\
$^{1}$Berkeley Center for Cosmological Physics, Berkeley, CA 94720, USA \\
$^{2}$Lawrence Berkeley National Laboratory, Berkeley, CA, 94720, USA \\
$^{3}$Department of Astronomy \& Astrophysics and Dunlap Institute, University of Toronto, Toronto, ON M5S 3H4, Canada \\
$^{4}$Department of Physics \& Astronomy, University of Waterloo, 200 University Avenue West, Waterloo, ON, N2L 3G1, Canada
}

\pubyear{2018}

\begin{document}
\label{firstpage}
\pagerange{\pageref{firstpage}--\pageref{lastpage}}
\maketitle

\begin{abstract}
We have used the first $\sim$3 years of 3.4$\mu$m (W1) and 4.6$\mu$m (W2) observations from the WISE and NEOWISE missions to create a full-sky set of time-resolved coadds. As a result of the WISE survey strategy, a typical sky location is visited every six months and is observed during $\gtrsim$12 exposures per visit, with these exposures spanning a $\sim$1 day time interval. We have stacked the exposures within such $\sim$1 day intervals to produce one coadd per band per visit -- that is, one coadd every six months at a given
position on the sky in each of W1 and W2. For most parts of the sky we have generated six epochal coadds per band, with one visit during the fully cryogenic WISE mission, one visit during NEOWISE, and then, after a 33 month gap, four more visits during the NEOWISE-Reactivation mission phase. These coadds are suitable for studying long-timescale mid-infrared variability and measuring motions to $\sim$1.3 magnitudes fainter than the single-exposure detection limit. In most sky regions, our coadds span a 5.5 year time period and therefore provide a $>$10$\times$ enhancement in time baseline relative to that available for the AllWISE catalog's apparent motion measurements. As such, the signature application of these new coadds is expected to be motion-based identification of relatively faint brown dwarfs, especially those cold enough to remain undetected by Gaia.





\end{abstract}



\section{Introduction}
\label{sec:intro}

The Wide-field Infrared Survey Explorer \citep[WISE;][]{wright10} has surveyed the entire sky at four mid-infrared wavelengths:
3.4$\mu$m (W1), 4.6$\mu$m (W2), 12$\mu$m (W3) and 22$\mu$m (W4). WISE represents a unique and highly valuable resource, because there is no funded successor slated to provide a next-generation, full-sky survey at similar wavelengths. Therefore, maximizing the scientific return of the WISE data is
critical to many branches of astronomy, especially considering WISE's potential for identifying rare/exotic objects suitable for follow-up with the \textit{James Webb Space Telescope} \citep[JWST;][]{jwst}.

WISE successfully completed a seven month full-sky survey in all four of its channels beginning in 2010 January. WISE continued surveying the sky
in its bluest three bands during 2010 August and September. After that, it performed the NEOWISE asteroid hunting mission until 2011 February \citep{neowise}, with only the W1 and W2 channels remaining operational. Following a 33 month hibernation period, the WISE instrument recommenced survey operations in 2013 December \citep{neowiser}. This post-hibernation mission is referred to as NEOWISE-Reactivation (NEOWISER). WISE continues to survey the sky, and has already completed approximately eight full W1/W2 sky passes over the course of its first seven years in orbit.



The NEOWISER mission is optimized for minor planet science and therefore publishes single-exposure images and catalogs, but delivers no coadded data products. The full archive of W1/W2 single-frame images now includes over
21 million exposures comprising $\sim$140 terabytes of pixel data. As such, reprocessing this archival data to create products tailored for science beyond the main belt represents a computationally formidable task. Nevertheless, the astronomy research community has recognized the great value in carrying out that work \citep[e.g.][]{faherty}.


We are undertaking an effort to repurpose the vast set of NEOWISER single-frame images for Galactic and extragalactic astrophysics by combining them with pre-hibernation imaging to build coadded data products. In \cite{meisner17} and \cite{meisner16} we combined WISE and NEOWISER observations to create the deepest ever full-sky maps at 3--5 microns. Those coadds are particularly
valuable for cosmology applications in which the most sensitive possible static sky maps are required \citep[e.g.][]{desi, desi_part1, desi_part2}. However, such `full-depth' coadds do not maximally capitalize on the extremely strong time-domain aspect of the combined WISE+NEOWISER data set.


The WISE survey strategy is such that, at a typical sky location, the available exposures can be segmented into a series of $\sim$1 day time intervals (referred to as visits), with such visits
occurring once every six months. In this work we present a full-sky set of time-resolved coadds which, at each location on the sky, stack together those exposures grouped within the same visit. Over most of the sky, this yields six coadd epochs spanning a 5.5 year time baseline. Because a visit typically entails $\gtrsim$12 exposures overlapping a particular sky location, our epochal coadds allow for detection of sources $\sim$1.3 mags fainter than the single-exposure depth. Thus, these coadds provide a powerful new tool for studying variability and motions of relatively faint WISE sources. The coadds are optimized for characterizing motion/variability on 
long timescales of 0.5--5.5 years, at the expense of averaging away variations on short ($\lesssim$1 day) timescales.

We construct our time-resolved coadds with two primary scientific applications in mind. The first is quasar variability. \cite{kozlowski10} currently represents the definitive
study of mid-infrared quasar variability, based on $\sim$3,000 spectroscopically confirmed AGN within $\sim$8 square degrees of Spitzer [3.6] and [4.5] repeat imaging. For comparison, our coadds will facilitate variability studies incorporating tens of thousands of mid-IR bright quasars spectroscopically confirmed by SDSS/BOSS over $\sim$1/3 of the sky \citep{sdss_quasars}.






The second application motivating our time-resolved coadds is motion-based discovery of cold, nearby substellar objects. Although WISE has already revolutionized the study of brown dwarfs \citep[e.g.][]{kirkpatrick11, cushing11}
and reshaped our knowledge of the immediate solar neighborhood \citep[e.g.][]{j1049, j0855, scholz14, mamajek15}, a large parameter space remains untapped. The surveys of \cite{luhman14}, 
\cite{allwise_motion}, \cite{schneider16} and \cite{allwise2_motion} have underscored the discovery potential of WISE-based brown dwarf searches which rely primarily on motion 
selection rather than color cuts. However, such searches with WISE have measured motions by grouping single-exposure detections and/or employing a short $\sim$6 month WISE time baseline, limiting them to relatively bright samples. With our epochal coadds, WISE-based brown dwarf motion searches can be extended fainter by more than a magnitude \citep[e.g.][]{kuchner17}.


In $\S$\ref{sec:data} we describe the archival data used in the course of this work. In $\S$\ref{sec:coaddition} we provide details of our coaddition methodology. In $\S$\ref{sec:results} we present and evaluate the coadd
outputs. In $\S$\ref{sec:dr} we provide information about our time-resolved coadd data release. We conclude in $\S$\ref{sec:conclusion}.


\section{Data}
\label{sec:data}

In this work we employ all of the publicly released W1/W2 Level 1b (L1b) single-exposure images with acquisition dates between 7 January 2010 and 13 December 2015 (UTC). In other words, we use all pre-hibernation W1/W2 L1b frames, plus those frames included in the first two annual NEOWISER data releases. In 2017 June, an additional year of 
W1/W2 exposures became public as part of the third annual NEOWISER data release. This most recent year of publicly available W1/W2 frames has not yet been incorporated into
the time-resolved coadds presented in this work, though we plan to publish updates with additional coadd epochs in the future.


We begin by downloading a local copy of all W1/W2 L1b sky images (\verb|-int-| files) along with their corresponding uncertainty maps (\verb|-unc-| files) and bitmasks (\verb|-msk-|
files). We refer to one such group of six files (three for each of W1 and W2) resulting from a single WISE pointing as a `frameset'. The coadds presented in this work make use of 7.9 million such framesets, for a total of 52 TB per band and 49 terapixels of single-exposure inputs.

We note that $\sim$50\% ($\sim$15\%) of the sky has sufficient W3 (W4) data to construct two epochs worth of time-resolved coadds in these bands. However, because of 
the partial sky coverage, small number of epochs, and short time baseline that would be associated with such W3/W4 time-resolved coadds, we have opted not to
create them.

\section{Coaddition Details}
\label{sec:coaddition}

Our time-resolved coadds are generated using an adaptation of the unWISE  \citep{lang14}  code. We describe our modifications of the unWISE coaddition pipeline in $\S$\ref{sec:updates}. Complete details of the unWISE coaddition
methodology are provided in \cite{lang14}, \cite{meisner17} and \cite{meisner16}. unWISE coadds serve as full-resolution alternatives to the WISE team's Atlas stacks \citep{cutri12, cutri13, masci09}. The latter are constructed using a point response function interpolation kernel that makes them optimal for detecting isolated sources but degrades their angular resolution. unWISE, in contrast, uses third order Lanczos interpolation to preserve the native WISE angular resolution. The unWISE coadds are optimized for forced 
photometry like that of \cite{lang14b}, where optical imaging with superior angular resolution provides source centroids and morphologies that are held fixed while determining best-fit WISE fluxes for each object. There are no full-sky Atlas data products comparable to the time-resolved coadds presented in this work, although it is possible to 
generate custom Atlas-like coadds for select sky regions and user-specified observation date ranges with IRSA's online `ICORE' WISE/NEOWISE coadder\footnote{\url{http://irsa.ipac.caltech.edu/applications/ICORE}}.

\subsection{Tiling}
\label{sec:tiling}

unWISE coadds adopt the same set of 18,240 tile centers as those used for the Atlas stacks. These tile centers trace out a series of iso-declination rings and their $x$/$y$ axes are oriented along the equatorial cardinal directions. Each
tile is 1.56$^{\circ}$$\times$$1.56^{\circ}$ in angular extent, with some overlap between neighboring tile
footprints ($\sim$3$'$ at each boundary at the equator). unWISE coadds employ a pixel scale matching that of the L1b images, 2.75$''$ per pixel, so that each coadd is 2048 pixels on a side. Each of the 18,240 unique astrometric footprints is identified by its \verb|coadd_id| value, a string encoding the tile's central
(RA, Dec) coordinates. For example, the tile centered at (RA, Dec) = (130.04$^{\circ}$, $-$18.17$^{\circ}$) has \verb|coadd_id| = 1300m182.



\subsection{Time-slicing}

\subsubsection{WISE Survey Strategy}
\label{sec:survey}
Before turning to the details of our time-slicing algorithm, it is helpful to review the WISE survey strategy, which motivates our choices of coadd epoch boundaries. WISE resides
in a Sun-synchronous low-Earth polar orbit. Early in the mission, WISE's orbit closely followed the Earth's terminator, and has gradually precessed off of the dawn-dusk boundary by $\sim$18$^{\circ}$ over the course of its 7$+$ years in orbit. WISE observes at nearly 90$^{\circ}$ solar elongation, pointing close to radially outward along the vector connecting the Earth's center and the spacecraft. In the course of one $\sim$95 minute orbit, WISE scans a great circle in ecliptic latitude at fixed ecliptic longitude. The ecliptic longitude of WISE scans progresses eastward at $\sim$1 degree per day. This rate of advancement in ecliptic longitude, in combination with WISE's 0.78$^{\circ}$$\times$0.78$^{\circ}$ field of view, means that a typical sky location at low ecliptic latitude will be observed during a series of exposures spanning $\sim$1 day each time its relevant sky region is visited. Another consequence of this survey strategy is that one full-sky mapping is completed every six months during which WISE is operational. Within one such six-month interval, half of the sky is scanned while WISE is
pointing `forward' along the direction of Earth's orbit, and the other half is scanned while WISE is pointing `backward' in the direction opposite Earth's orbit. Forward 
scans are those which scan from ecliptic north to ecliptic south, while backward scans progress from ecliptic south to ecliptic north. Putting this
all together, a given sky location is observed for a time period of order one day once every six months, with such visits alternating in scan direction.  A final consequence of the
WISE survey strategy is that the ecliptic poles are observed during nearly every scan. The ecliptic poles therefore require special treatment within our time-slicing procedure.


\subsubsection{Defining Epoch Boundaries}
\label{sec:slicing}

Here we describe the algorithm by which exposures relevant to a given \verb|coadd_id| astrometric footprint are segmented into a series of coadd epochs. For each \verb|coadd_id|,
our time-slicing procedure is applied independently in each WISE band. We begin by identifying the set of all exposures in the band of interest with central coordinates close enough to the \verb|coadd_id| tile center to contribute (angular separation less than $1.66^{\circ}$). We then sort these exposures by MJD. Next, we loop over the sorted exposures, starting with the earliest, and define an epoch boundary any time that a gap of $>$90 days is found between consecutive exposures. Such $>$90 day gaps are taken to be indicative of the periods between six-monthly visits.

Typically, at low ecliptic latitude, the exposures grouped into a `time slice' between two consecutive epoch boundaries span a period of order a few days. However, close to the ecliptic poles, where WISE obtains nearly continuous coverage, this simplistic picture of several-day time slices spaced at six-monthly intervals breaks down. As a result, the procedure outlined above yields a small number of very long time slices which can last months or years. This is not desirable from the perspective of measuring variability near the ecliptic poles. Therefore, for \verb|coadd_id| footprints centered at $|\beta| > 80^{\circ}$ ($\beta$ denotes ecliptic latitude), we subdivide the time slices with exposures that span more than 15 days into a series of shorter epochs, each of which lasts 10 days or less. We accomplish this by looping over the MJD-sorted exposures within each $>$15 day time slice, creating a new epoch boundary any time that an exposure is encountered which was acquired more than 10 days after the exposure immediately following the previous epoch boundary.

Each time slice we create is identified by a unique (\verb|coadd_id|, \verb|band|, \verb|epoch|) triplet. \verb|band| is simply an integer, 1 (2) for W1 (W2). \verb|epoch| is a zero-indexed
integer counter indicating the time slice's temporal ordering relative to other time slices of the same (\verb|coadd_id|, \verb|band|) pair. The only rules in assigning the \verb|epoch| number to time slices are:

\begin{enumerate}
\item \verb|epoch| always starts at 0 for the earliest time slice of each (\verb|coadd_id|, \verb|band|) pair.
\item \verb|epoch| always increases with time for a given (\verb|coadd_id|, \verb|band|) pair, and is incremented by 1 for each successive time slice.
\end{enumerate}


This convention has one particularly notable implication: it is not safe to assume that coadd images with the same \verb|epoch| value but differing \verb|coadd_id| values have similar MJDs. For instance, most \verb|coadd_id| values will have two available pre-hibernation visits, so that \verb|epoch| = 2 corresponds to the first post-reactivation
visit. However, $\sim$1/6 of the sky was visited a third time in early 2011, and for the affected \verb|coadd_id| footprints, \verb|epoch| = 2 happens \textit{before} hibernation. One could imagine an alternative convention in which each unique \verb|epoch| value corresponds by definition to a particular range of MJD across all \verb|coadd_id| values, but this is not what we have chosen to do.

We assign \verb|epoch| values during the process of determining coadd epoch boundaries, which occurs prior to the actual coaddition. In rare cases, this can lead to situations
where a particular (\verb|coadd_id|, \verb|band|) pair ends up with coadd outputs for \verb|epoch| = $(N+1)$ but not \verb|epoch| = $N$, with $N \ge 0$. This can happen, for example, when all of the frames slated for inclusion in \verb|epoch| = $N$ turn out to narrowly avoid overlapping the relevant \verb|coadd_id| astrometric footprint.



\subsection{unWISE Pipeline Updates Relative to Full-depth Coaddition}
\label{sec:updates}

The primary difference in our coaddition methodology relative to previous unWISE releases is our segmentation of L1b exposures according to the time-slicing rules
of $\S$\ref{sec:slicing}, whereas past unWISE stacks simply added all exposures together regardless of their MJDs.

A second substantial difference between our time-resolved coaddition and previous unWISE processings is that the epochal coadds described in this work have been built using full-depth unWISE stacks as reference templates. In \cite{meisner16}, we introduced a framework for recovering moonlight-contaminated WISE frames by comparing them to an appropriate Moon-free reference stack, thereby deriving low-order polynomial corrections which removed time-dependent background variations in problematic L1b images. During our time-resolved coaddition procedure,
we apply this same low-order polynomial correction methodology to \textit{every} L1b exposure. We find that these corrections help to ensure that there are no edges of L1b exposures imprinted at low levels in our time-resolved coadds. We use the \cite{meisner17} full-depth coadds as reference templates. In detail, the polynomial background corrections are fit and applied on a per-quadrant basis for each L1b exposure. Within each quadrant, a fourth order polynomial correction is computed relative to the full-depth stack and subtracted from the corresponding resampled L1b pixels.

\section{Results}
\label{sec:results}

\begin{figure}
        \includegraphics[width=3.3in]{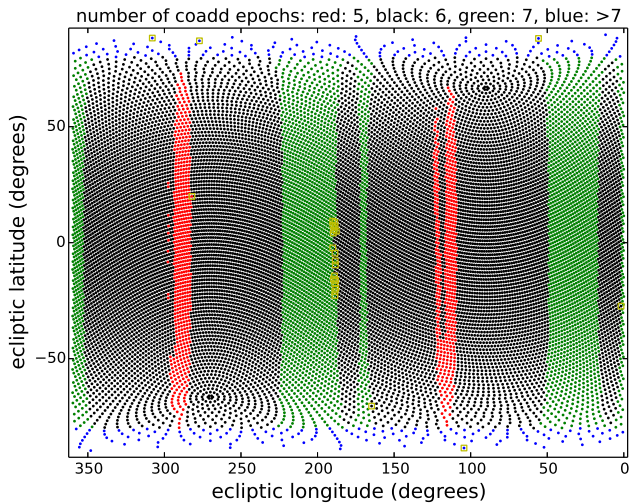}
    \caption{Number of W1 coadd epochs per \texttt{coadd\_id} astrometric footprint, shown in ecliptic coordinates. Yellow boxes denote footprints for which the number of W1 coadd epochs differs from the number of W2 coadd epochs. Typically there are six coadd epochs available per band (black dots). The red-colored ranges of ecliptic longitude have only five coadd epochs per band
because they were impacted by the command timing anomaly in 2014 April. The green-colored ranges of ecliptic longitude have an extra seventh epoch because they were
observed during the partially complete third sky pass of the pre-hibernation mission. The ecliptic poles ($|\beta|$ $>$ 80$^{\circ}$) are blue, indicating $>$7 available coadd epochs in each band as a result of the modified time-slicing rules employed for these regions (see $\S$\ref{sec:slicing}).}
    \label{fig:num_epochs}
\end{figure}

In total we have generated 234,740 single-band sets of time-resolved unWISE coadd outputs. Figure \ref{fig:num_epochs} shows a map of the number of available coadd
epochs in ecliptic coordinates. Typically there are six coadd epochs available per band per \verb|coadd_id| footprint. The maximum number of epochs for any (\verb|coadd_id|, \verb|band|) pair is 112, which happens near $|\beta|$ = 90$^{\circ}$. The minimum number of coadd epochs for any (\verb|coadd_id|, \verb|band|) pair is 5.


\begin{figure*}
       \includegraphics[width=7.0in]{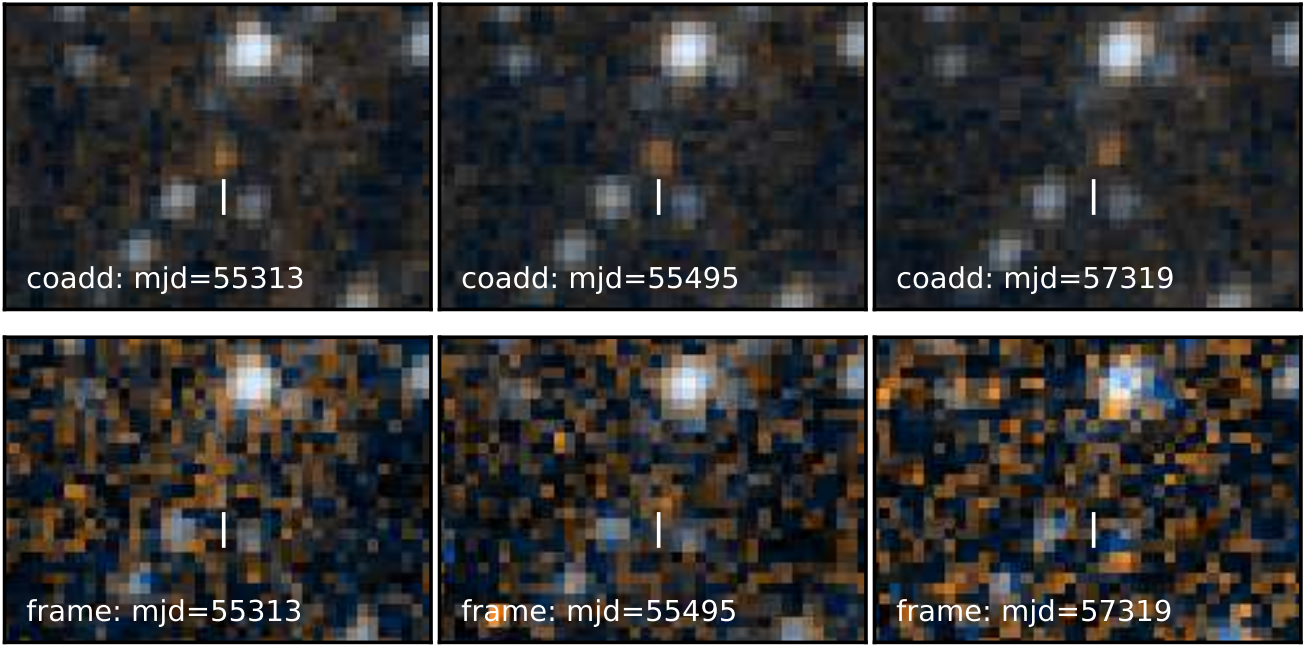}
   \caption{The T6.5pec brown dwarf WISE J201404.13+042408.5 \citep{mace13}, with W2 $\approx$ 15.0 and W1 $\approx$ 17.3 according to the AllWISE catalog. Top left: Color composite of \texttt{epoch} = 0 of \texttt{coadd\_id} = 3037p045, centered on the brown dwarf's location. The field of view is $1.8'$ $\times$ $1.3'$. The colormap's blue (orange) channel represents W1 (W2). Top center: Same, but for \texttt{epoch} = 1, the final visit prior to hibernation, which occurs $\sim$0.5 years after the first coadd epoch. Top right: Same, but for \texttt{epoch} = 5, the final visit available in our time-resolved coadd data release. The white line remains fixed in all panels to accentuate the brown dwarf's motion. A hint of motion may be discernible between the two pre-hibernation coadd epochs spanning a $\sim$0.5 year time period, whereas the positional shift between the first and final coadd epochs is visually obvious, highlighting the importance of our coadds' $\sim$5.5 year time baseline. At $\sim$0.5 magnitudes below the single-exposure detection limit, this brown dwarf is effectively imperceptible in individual frames. Each panel in the bottom row displays a randomly selected frame which contributes to the coadd shown directly above it, illustrating the importance of coaddition in observing this brown dwarf's
   motion with WISE. With a predicted optical magnitude of $G \approx $ 24.6 \citep{smart17}, this brown dwarf is expected to be far below the Gaia detection limit.}
    \label{fig:bd}
\end{figure*}

\begin{figure*}
       \includegraphics[width=7.0in]{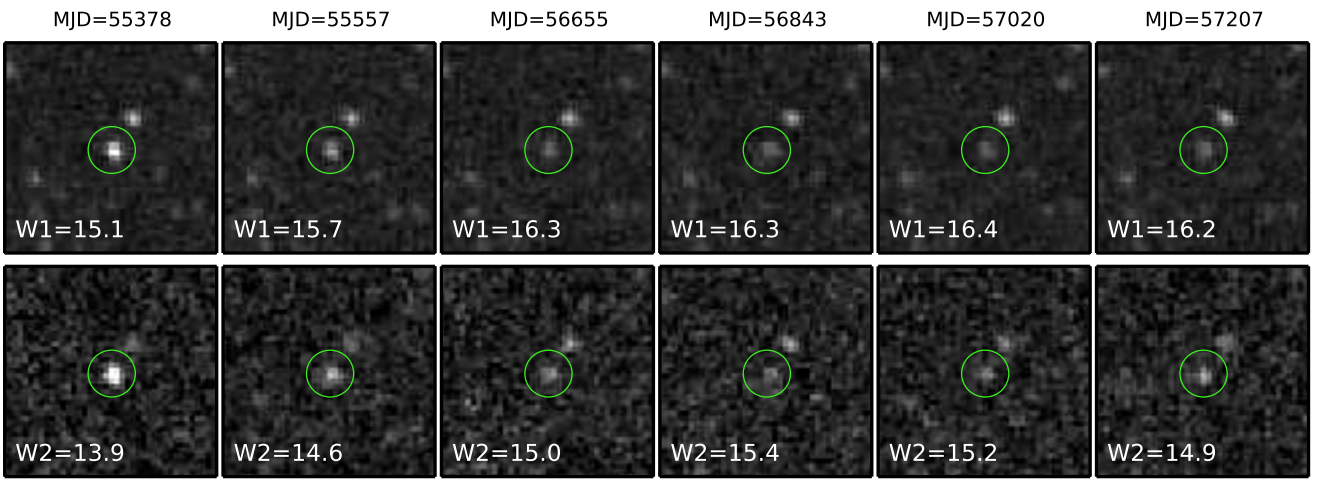}
   \caption{SDSS J001539.99+170040.5, an AGN from the SDSS quasar catalog \citep{sdss_quasars}. Top row: Cutouts drawn from our six available W1 coadd epochs. Bottom row: Cutouts drawn from our six available W2 coadd epochs. The field of view is $1.9'$$\times$$1.9'$ and magnitudes are in the Vega system. Fluxes are measured via \textit{Tractor} forced photometry \citep{lang14b}. The object fades by $>$1 mag in both W1 and W2 over the course of $\sim$4 years, then begins to brighten again. Importantly, in both W1 and W2, the object transitions from brighter
   than the single-exposure detection limit (W1=15.3, W2=14.5) to fainter than this threshold, which would complicate studies of such variability based on the
   L1b source database.}
    \label{fig:quasar}
\end{figure*}

Figure \ref{fig:bd} shows an example of a faint brown dwarf's motion as seen in our coadds. Figure \ref{fig:quasar} shows an example of a highly variable AGN as seen in our time-resolved coadds.

\subsection{Astrometry}
\label{sec:astrom}

Because motion-based brown dwarf discovery represents a key application of our time-resolved coadds, we believe it is important to present an extensive characterization/discussion
of the coadd astrometry. When resampling the L1b pixels to coadd space during coaddition, we simply take the WCS information in the L1b headers at face value -- the intention is to preserve the L1b astrometry, and we therefore do not perform any custom WCS recalibration or tweaking during coaddition.

\begin{figure*}
       \includegraphics[width=7.0in]{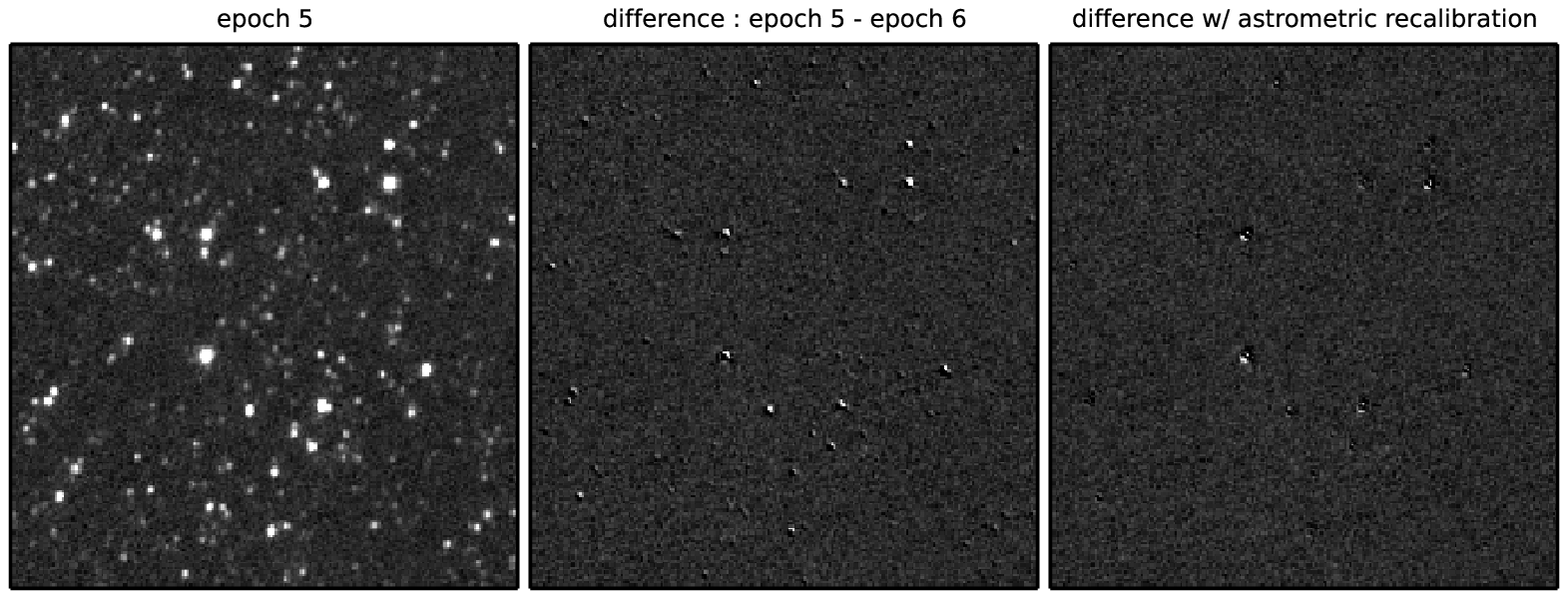}
   \caption{Example of the impact of our astrometric recalibration on difference image residuals. Left: 9.5$'$ $\times$ 10.2$'$ cutout from our epochal 
                coadd with \texttt{coadd\_id} = 2246p106,  
                 \texttt{band} = 1, \texttt{epoch} = 5, centered at ($\alpha$, $\delta$) = (224.93021$^{\circ}$, 10.88$^{\circ}$). Center: subtraction of the \texttt{epoch} = 6 coadd of the same \texttt{coadd\_id} in W1 from the \texttt{epoch} = 5 coadd, using the L1b-based WCS without recalibration. Right: same subtraction, but using the astrometric recalibration described in $\S$\ref{sec:scamp} to align the two epochal coadds. The dipoles in the center panel
                 are eliminated, leaving behind relatively high-order, small-amplitude residuals.}
    \label{fig:dipole}
\end{figure*}

\begin{figure*}
       \includegraphics[width=7.0in]{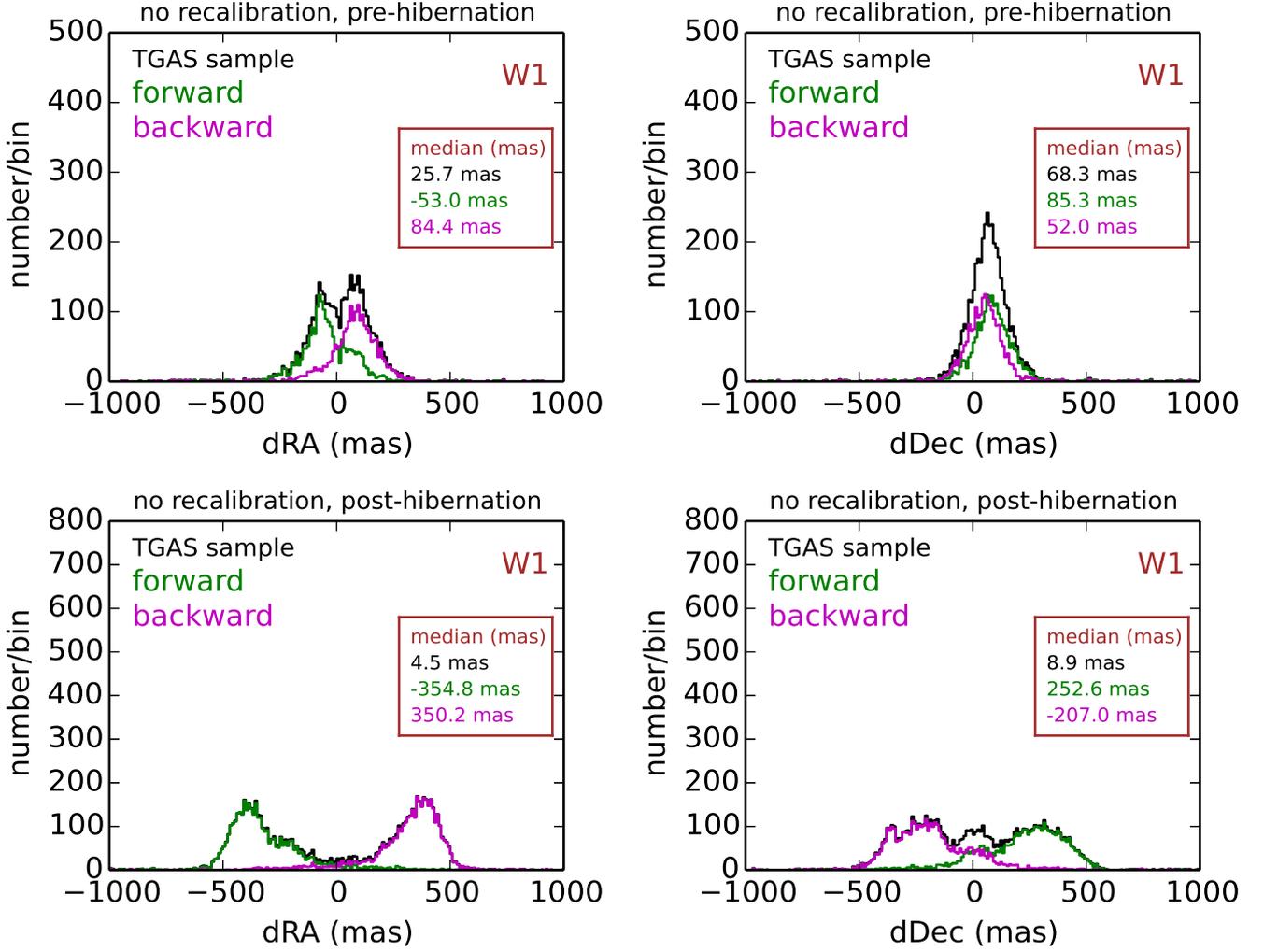}
   \caption{Astrometric residuals of our TGAS sample, converting our flux-weighted coadd centroid measurements to world coordinates using the L1b-based WCS adopted during coaddition. Top left: W1 pre-hibernation residuals in RA. Top right: W1 pre-hibernation residuals in Dec. Bottom left: W1 post-hibernation residuals
   in RA. Bottom right: W1 post-hibernation residuals in Dec. Residuals from forward-pointing (backward-pointing) scans are always shown as green (magenta) histograms, while
   the black histograms combine both scan directions together. The residuals shown in Figures \ref{fig:tgas_w1_raw}--\ref{fig:icrf_w2_recalib} are always measured positions
   minus `true' positions, with true positions coming from either TGAS or ICRF2 as indicated in the plot annotations.  RA residuals are in true angular units, i.e. we have multiplied the difference in RA by cos(Dec). The forward/backward sub-distributions which appear as roughly mirror images of each other result from the rotational asymmetry of the PSF models used to calibrate the L1b WCS, as explained in $\S$\ref{sec:astrom}. There are two reasons to segment the residuals by time into pre and post hibernation subplots: (1) The PSFs used to calibrate the L1b WCS are different during these two time periods. In detail, different PSF models were also used during different phases of the pre-hibernation mission, despite our lumping the entire pre-hibernation timespan together in these plots. (2) The L1b WCS calibration procedure ignored proper motion of the calibrators for pre-hibernation exposures, then
   began taking this effect into account for all post-hibernation data. This explains why the Dec residuals at upper right are shifted relative to zero.}
    \label{fig:tgas_w1_raw}
\end{figure*}

\begin{figure*}
       \includegraphics[width=7.0in]{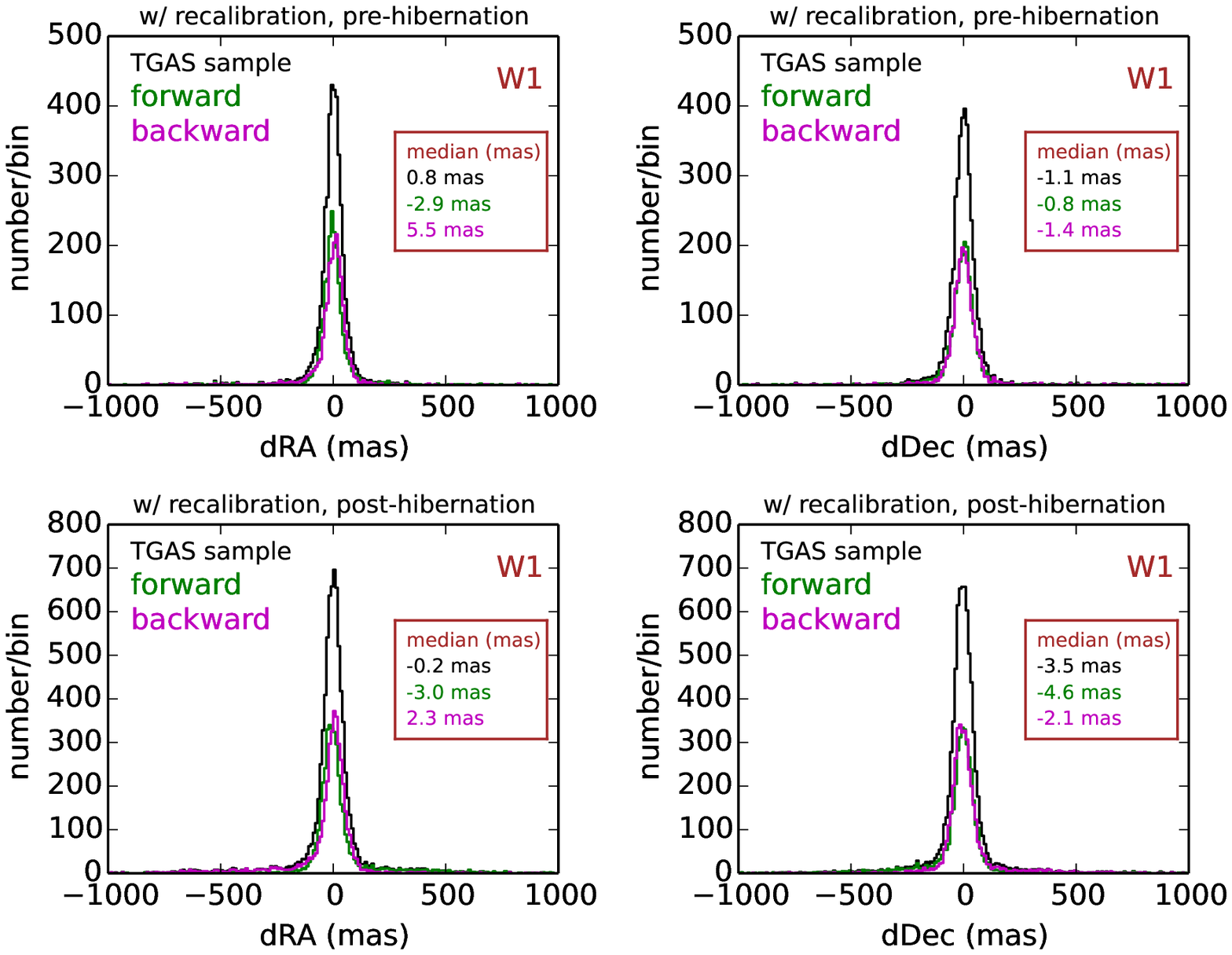}
   \caption{Same as Figure \ref{fig:tgas_w1_raw}, but showing residuals of our recalibrated W1 source positions relative to TGAS. The distributions of residuals, even when segmented by
   scan direction, are always sharply/singly peaked and centered very close to zero, indicating that our astrometric recalibration procedure of $\S$\ref{sec:scamp} has been highly successful for these normal-colored sources.}
    \label{fig:tgas_w1_recalib}
\end{figure*}

We have found that subtracting pairs of coadd epochs with the same \verb|coadd_id| and \verb|band| often reveals strong dipole residuals suggestive of astrometric misalignments
(see Figure \ref{fig:dipole} for an example). These misalignments are most pronounced when differencing pairs of coadd epochs which are built from exposures with opposite scan directions. As discussed in $\S$\ref{sec:survey}, at a given sky location not nearby the ecliptic poles, there will be two distinct (and roughly opposite) scan directions, which
correspond to scans pointing forward/backward along Earth's orbit. For convenience, we therefore refer to scans as either `forward-pointing' or `backward-pointing' throughout our astrometry analysis. Because of these two opposite scan directions, the coadd epochs of each \verb|coadd_id| can be thought of as having two corresponding `parities'. As a result of the WISE survey geometry, coadd epochs of a particular \verb|coadd_id| with the same parity are based on observations acquired from roughly the same point within Earth's orbit, and are
spaced at integer year intervals. On the other hand, coadd epochs of a single \verb|coadd_id| with opposite parity are acquired when Earth is on opposite sides of the Sun.

The explanation for dipole residuals like those seen in Figure \ref{fig:dipole} is that the PSF models adopted by the WISE team are not symmetric with respect
to swapping the scan direction (which, at a given sky location, corresponds to rotating the PSF model by roughly 180$^{\circ}$). The asymmetry is not merely present at low levels in the PSF model wings -- the flux-weighted PSF model centroid shifts by hundreds of milliarcseconds (mas) as a result of 180$^{\circ}$ rotation. In other words, the PSF models are substantially ``off-center" relative to their flux-weighted centroids\footnote{This feature of the WISE team's PSF modeling is documented in item III.2.m of the NEOWISE Explanatory Supplement's cautionary notes \citep{cutri15}.}. This has major implications for the L1b WCS solutions because these
were computed using source centroids derived from fits employing the WISE team's PSF models. The PSF model and astrometry are inherently degenerate, and because
the WISE team's PSF model centroid definition is substantially different from the flux-weighted centroid, we should expect to see that coadd images with opposite scan
directions appear shifted relative to one another when relying on the L1b WCS solutions during coaddition.


For concreteness, it is beneficial to examine a worked example. We adopt \verb|coadd_id|  = 0899p075, \verb|band| = 1 as a representative set of single-band epochal coadd outputs. \verb|epoch| = 2 (3) is the first (second) post-reactivation coadd epoch of this tile, and is built from backward-pointing (forward-pointing) scans. Consider a single extragalactic source within the \verb|coadd_id|  = 0899p075 footprint -- an object that can safely be deemed static due to its negligible parallax and proper motion. Because the WISE team's W1 PSF model centroid is offset from the flux-weighted centroid by $-$0.11 pixels along the detector $x$-axis, the flux-weighted centroid of this source is always shifted by $+$0.11 L1b pixels in the $\hat{x}$ direction relative to the location of this source according to the L1b WCS. The sign of this shift in detector coordinates is independent of scan direction. During forward-pointing scans the 
 $+\hat{x}$ direction points toward ecliptic west, while the $+\hat{x}$ direction points toward ecliptic east during backward-pointing scans, as the detector has undergone a 180$^{\circ}$ rotation between the forward/backward pointing scans.  Therefore, the flux-weighted source centroid measured in (\verb|coadd_id| = 0899p075, \verb|epoch| = 2, \verb|band| = 1) will be offset by 2$\times$(0.11 L1b pixels) = 0.6$''$ to the ecliptic east, relative to the same source's flux-weighted centroid in (\verb|coadd_id| = 0899p075, \verb|epoch| = 3, \verb|band| = 1). This gives rise to dipole residuals like those seen in the center panel of Figure \ref{fig:dipole}.
 
Generalizing this example, flux-weighted centroids \textit{in world coordinates} will be shifted equally in \textit{opposite} directions in coadd epochs of opposite  parities, relative to the source's sky position according to the L1b-based WCS. Therefore, when measuring flux-weighted centroids from our epochal coadds and converting them to world coordinates according to the L1b-based WCS, we should expect to see per-coordinate residuals relative to ground truth which show bimodal distributions. Each distribution of per-coordinate astrometric residuals should have two peaks placed symmetrically about zero, one peak corresponding to forward-pointing coadd epochs and the other to backward-pointing coadd epochs. Indeed, this qualitative scenario is seen in Figures \ref{fig:tgas_w1_raw}, \ref{fig:tgas_w2_raw}, \ref{fig:icrf_w1_raw} and \ref{fig:icrf_w2_raw}. This behavior ultimately stems from the inconsistency between coadd-level flux-weighted centroiding and the centroiding used to calibrate the L1b WCS.

One should expect to avoid systematics dependent on scan direction by measuring centroids with the same PSF models used to centroid L1b astrometric calibrators. Indeed, the L1b astrometric solutions and centroid measurements in the WISE team's NEOWISE-R Single Exposure Source Table are exquisitely self-consistent, at the level of a few mas or better according to the top panel of Figure \ref{fig:l1b_shifts}. So one solution to the relative shifts between coadd images of opposite parities is simply to \textit{always} measure source
positions using a PSF model that is consistent with the PSF model used to centroid the L1b WCS calibrators. Unfortunately, it would require substantial 
effort to translate exposure-level WISE PSF models into coadd-level PSF models corresponding to each of our epochal coadds. In detail there is not even a unique coadd-level PSF per epochal coadd because different sets of input exposures contribute to different sub-regions of each stacked image.

We anticipate that most users of our coadds will not attempt to build such PSF models, but rather use standard software packages \citep[e.g. Source Extractor;][]{sextractor} to
compute flux-weighted centroids and/or perform difference imaging \citep[e.g. with SWARP;][]{terapix}. In both of these applications, it is valuable to have `recalibrated' astrometry
for each epochal coadd, where the WCS calibration has been performed using flux-weighted centroids. We therefore have attempted to compute
such recalibrated astrometry for all of our epochal coadds. The following subsections discuss details of this recalibration and our evaluation of the recalibrated astrometric 
residuals relative to external data sets.

\subsection{Cataloging Source Centroids}
\label{sec:se}
The first step towards performing our astrometric recalibrations is deriving catalogs which include flux-weighted centroids measured within each of our epochal coadds. We do so using
Source Extractor \citep{sextractor}. We set a relatively high threshold for source detection, since we only seek sufficient numbers of bright sources to perform and evaluate our
astrometric recalibrations. The signal-to-noise distribution of our extractions typically peaks at $\sim$8.5, based on the \verb|FLUX_AUTO| and \verb|FLUXERR_AUTO| parameters. We adopt the \verb|XWIN_IMAGE|, \verb|YWIN_IMAGE| Source Extractor outputs for our centroid measurements. These are `windowed' flux-weighted centroids, meaning that the per-pixel weights used for standard isophotal flux weighting are tapered by a symmetric Gaussian envelope with FWHM similar to that of the PSF. This windowing results in reduced scatter without introducing bias. Running Source Extractor on all 234,740 of our epochal coadds yields catalogs of $\sim$1.68 billion (0.93 billion) extractions in W1 (W2). 

\subsection{Fitting Recalibrated WCS Solutions}
\label{sec:scamp}

We use SCAMP \citep{scamp} to fit WCS solutions based on our flux-weighted Source Extractor centroids. We must select a catalog of astrometric calibrators with respect to which SCAMP will solve for the WCS parameters. We have opted to use the HSOY catalog \citep{hsoy}, a Gaia DR1 \citep{gaia_dr1} plus PPMXL \citep{ppmxl} hybrid. We chose HSOY because it is a full-sky catalog that allows us to accurately account for the proper motion of each calibrator source, with typical per-coordinate uncertainties of $\sim$10 mas over the range of MJDs relevant to WISE observations.

For each Source Extractor catalog from $\S$\ref{sec:se}, we compute a corresponding HSOY catalog with positions translated to the appropriate MJD of the relevant coadd epoch by accounting for proper motion. No attempt is made to correct for parallax on a per-object basis, since the vast majority of HSOY sources have no Gaia parallax measurement available. We then perform a first order WCS solution with SCAMP for every (\verb|coadd_id|, \verb|epoch|, \verb|band|) triplet. The median number of sources used to compute the SCAMP solutions
is $\sim$3,100 (2,100) in W1 (W2) per epochal coadd at high Galactic latitude ($|b_{gal}| > 30^{\circ}$). The calibrators have median magnitudes of 13.7 Vega in both W1 and W2. The median color of our calibrator sources is (W1$-$W2) = $-0.04$, which is similar to the color of typical stars.

The SCAMP output parameters ASTRRMS1 and ASTRRMS2 provide per-coordinate, per-coadd measurements of residual astrometric scatter relative to the calibrators, restricted to bright (S/N $>$ 100), unsaturated sources. Both parameters have median values of 92 mas in high Galactic latitude regions ($|b_{gal}| > 30^{\circ}$), which corresponds to $\sim$1/30 of a WISE L1b pixel. We ran several thousand test cases worth of
SCAMP solutions with higher-order \verb|PV |distortions, but found that doing so led to a negligible decrease in the typical residual astrometric scatter. We therefore deemed higher-order distortion terms unnecessary. Our recalibrated astrometry is relative, not absolute, as we make no attempt to correct for the average parallax of our calibrator sources.


\subsection{Evaluation of Recalibrated Astrometry}

In this subsection we seek to answer the following question: what is the systematics floor of our recalibrated astrometry? We proceed by analyzing
two distinct sets of calibrators for which we have astrometric `ground truth'. The first sample is a subset of $\sim$1,900 Gaia TGAS sources \citep{tgas}. The second is 
a sample of $\sim$2,100 ICRF2 quasars \citep{icrf2}.

\begin{figure*}
       \includegraphics[width=7.0in]{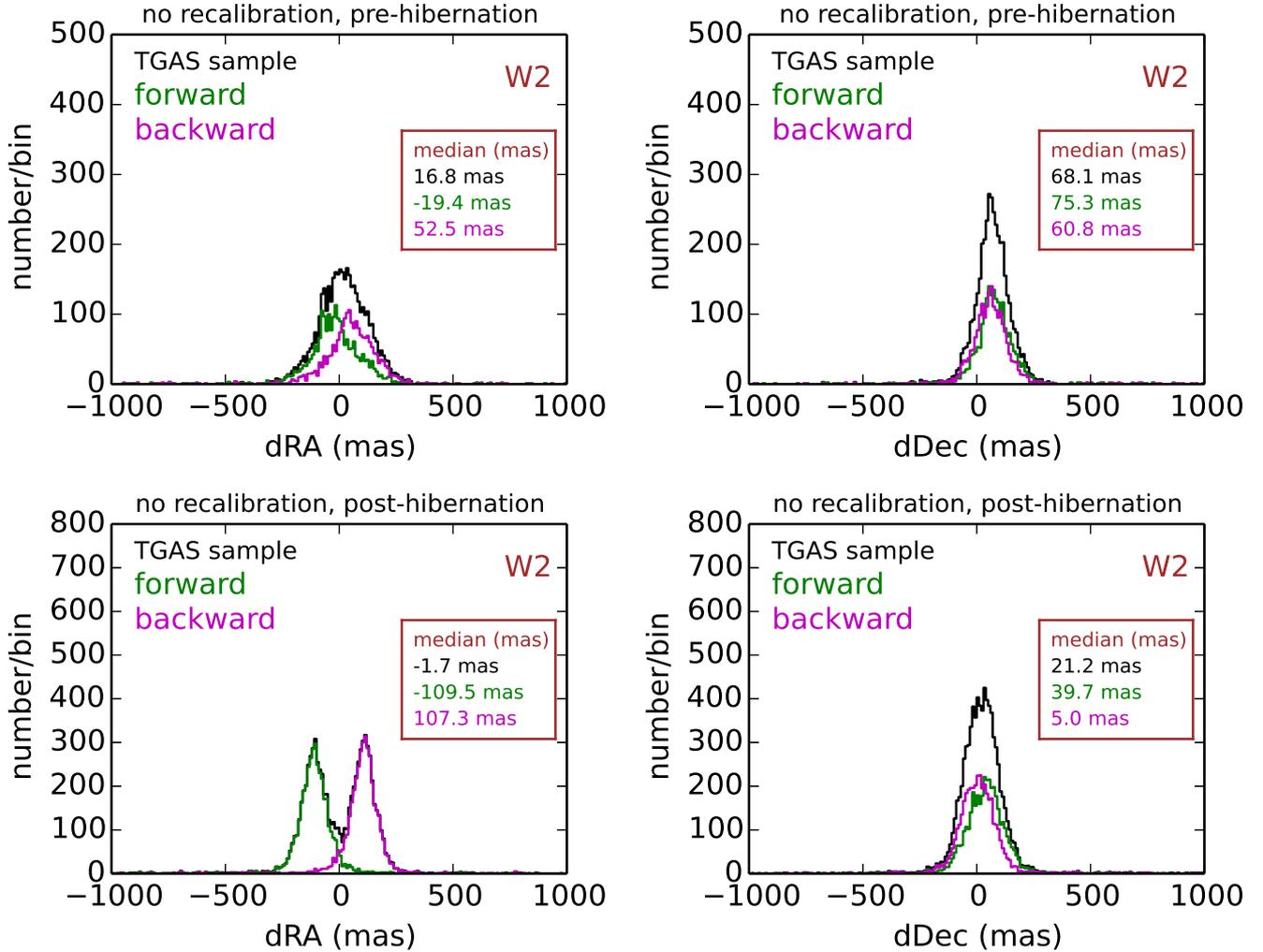}
   \caption{Same as Figure \ref{fig:tgas_w1_raw}, but for W2 rather than W1. Without astrometric recalibration, there are distinct sub-distributions corresponding to forward/backward scans, especially in the RA direction.}
    \label{fig:tgas_w2_raw}
\end{figure*}

\begin{figure*}
       \includegraphics[width=7.0in]{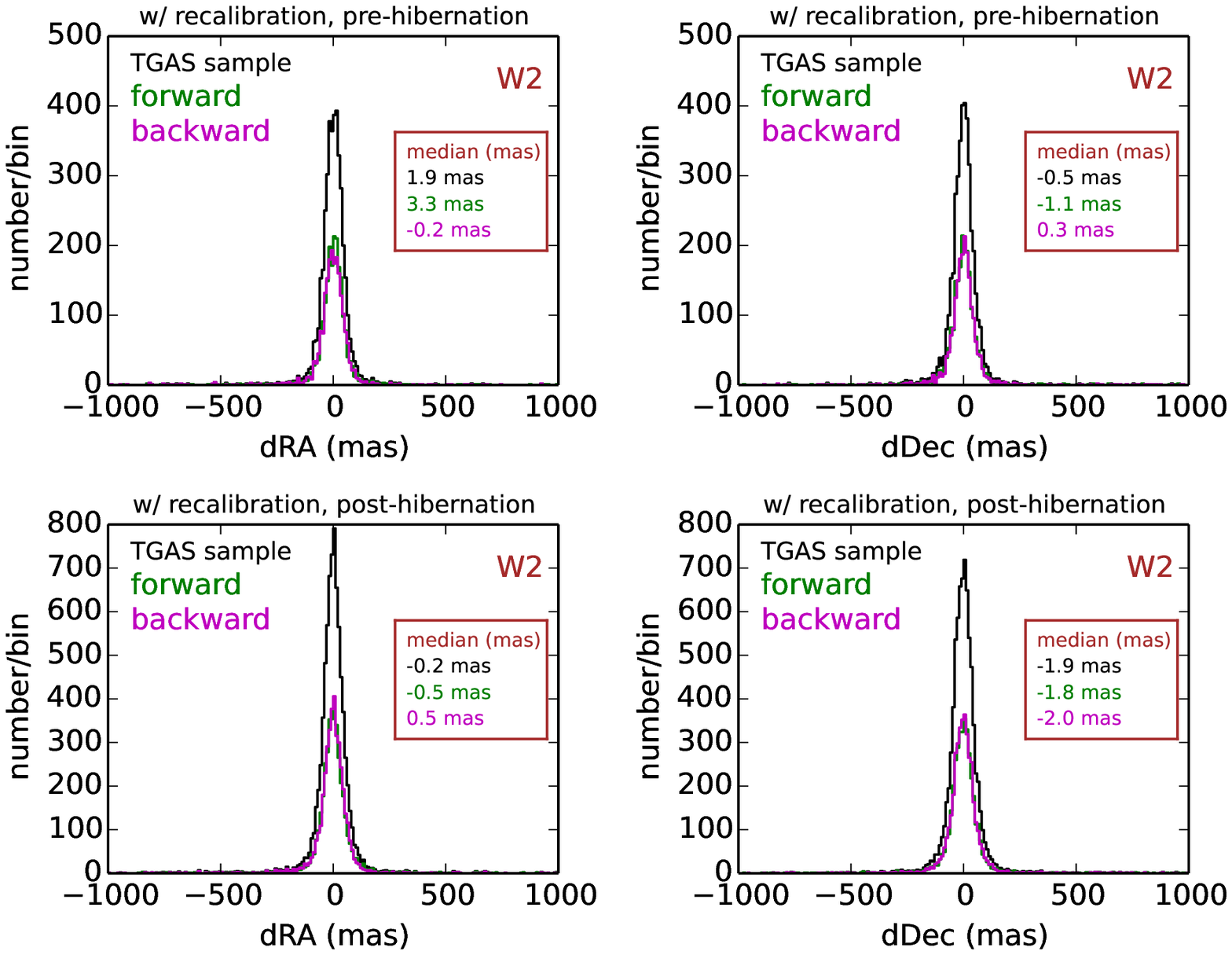}
   \caption{Same as Figure \ref{fig:tgas_w2_raw}, but showing residuals of our recalibrated W2 source positions relative to TGAS. The distributions of residuals, even when segmented by
   scan direction, are always sharply/singly peaked and centered very close to zero, indicating that our astrometric recalibration procedure of $\S$\ref{sec:scamp} has been highly successful for these normal-colored sources.}
    \label{fig:tgas_w2_recalib}
\end{figure*}

\begin{figure*}
       \includegraphics[width=7.0in]{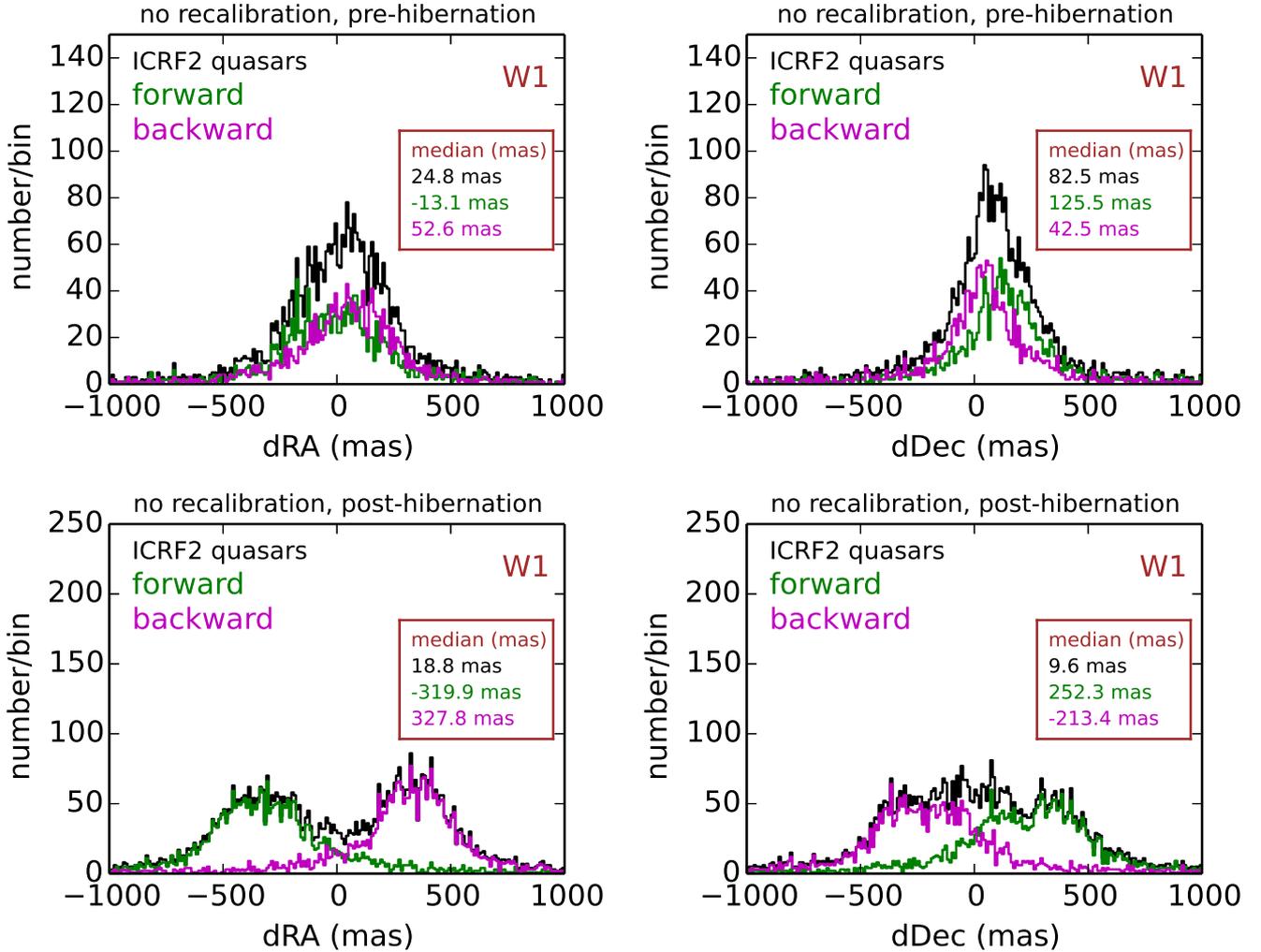}
   \caption{Same as Figure \ref{fig:tgas_w1_raw}, but for our ICRF2 quasar sample rather than our TGAS sample. The distributions are much broader than in Figure \ref{fig:tgas_w1_raw} because our ICRF2 sample is several magnitudes fainter than our TGAS sample at W1, leading to increased statistical scatter in our measured centroids.}
    \label{fig:icrf_w1_raw}
\end{figure*}

\begin{figure*}
       \includegraphics[width=7.0in]{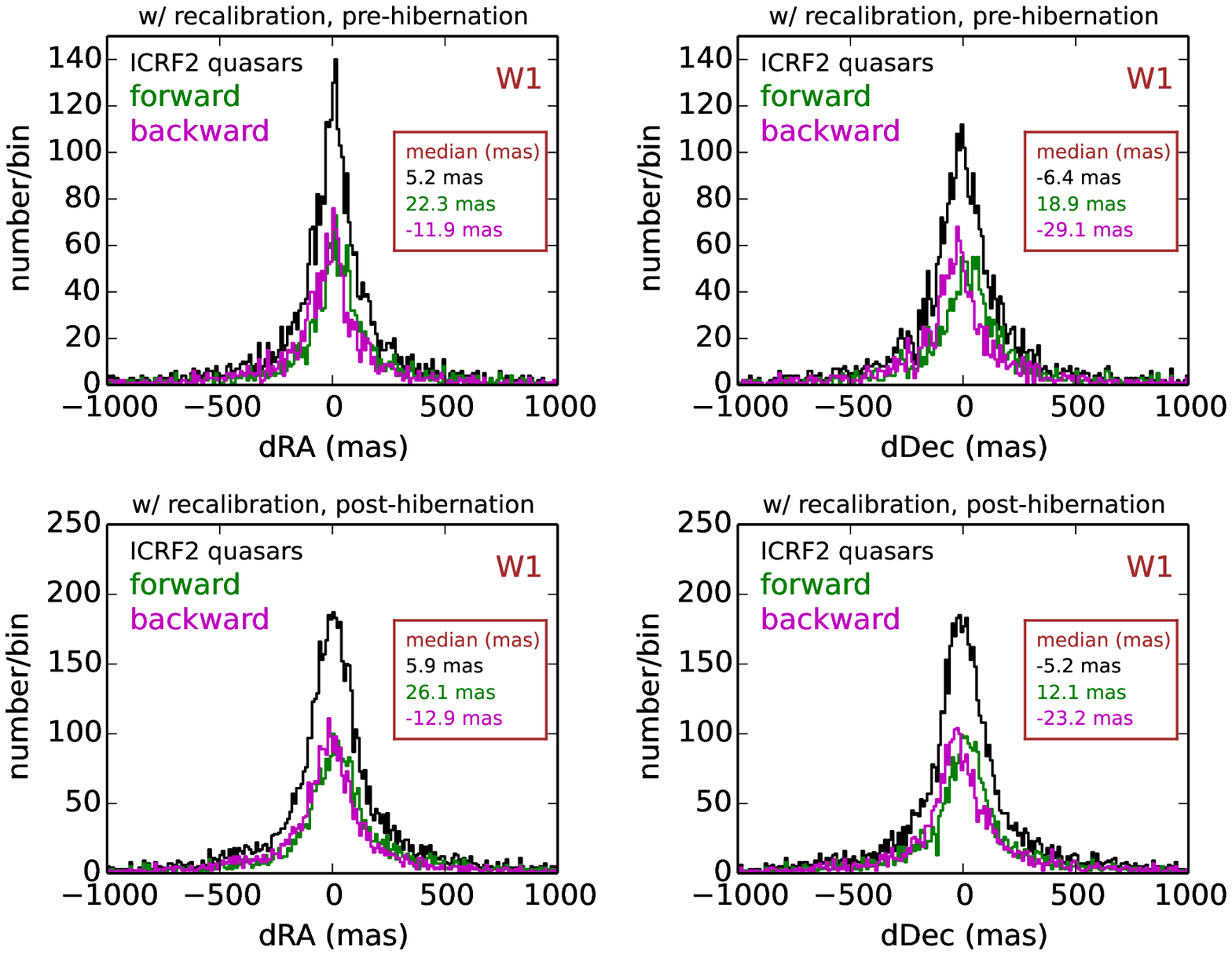}
   \caption{Same as Figure \ref{fig:icrf_w1_raw}, but showing residuals of our recalibrated ICRF2 source positions. The black distributions are sharper than in Figure \ref{fig:icrf_w1_raw}, but the forward-pointing/backward-pointing sub-distributions display some remaining separation, which we suggest may be due to a chromatic astrometric
   bias affecting these relatively red quasars ($\S$\ref{sec:chromatic}). }
    \label{fig:icrf_w1_recalib}
\end{figure*}

\begin{figure*}
       \includegraphics[width=7.0in]{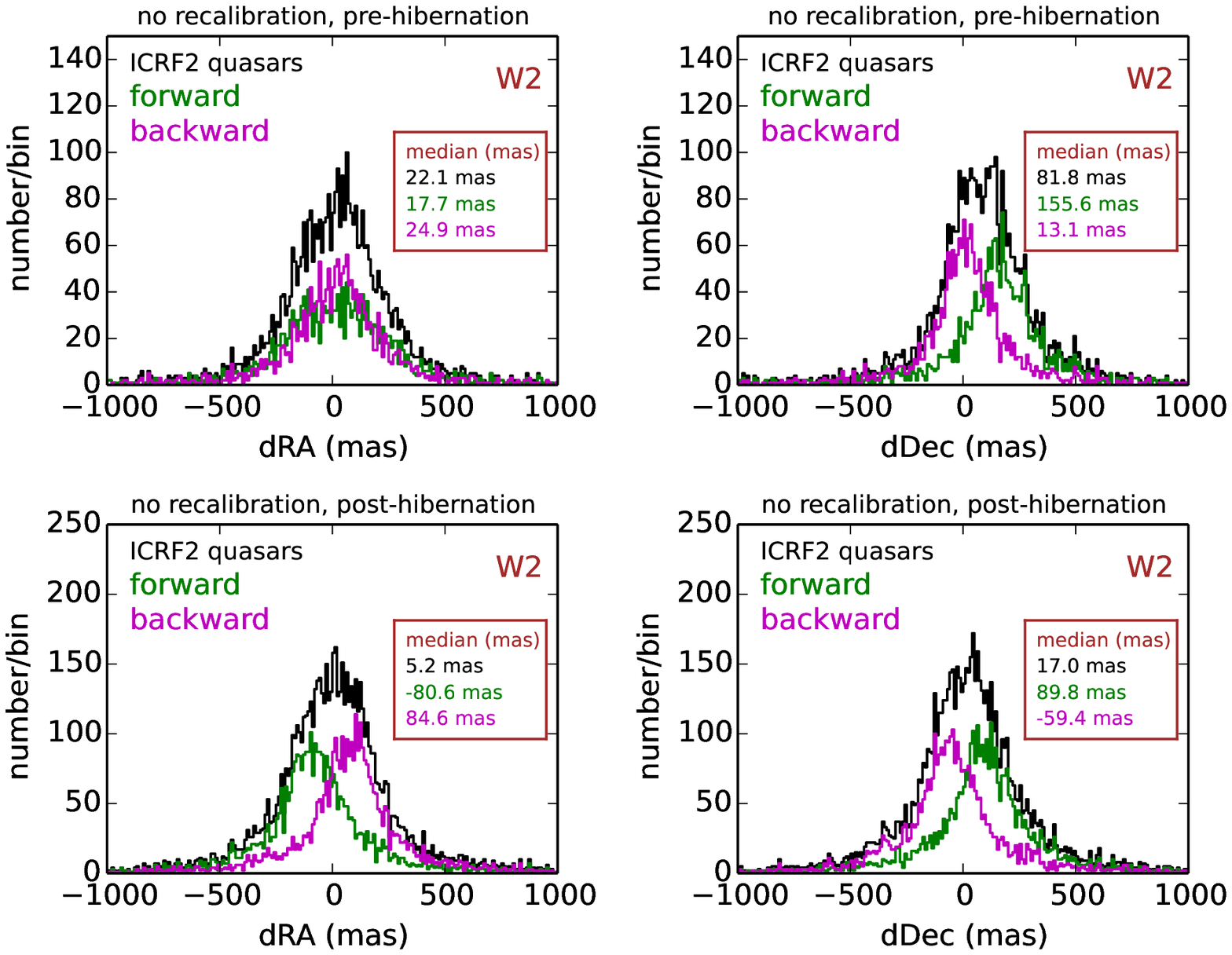}
   \caption{Same as Figure \ref{fig:tgas_w1_raw}, but for our ICRF2 quasar sample rather than our TGAS sample and in W2 rather than W1.}
    \label{fig:icrf_w2_raw}
\end{figure*}

\subsubsection{TGAS}
\label{sec:tgas}
Several considerations went into selecting a TGAS sample with which to evaluate our recalibrated astrometry. In particular, we sought out relatively faint TGAS sources, since
many TGAS objects are so bright as to saturate in W1/W2. We also desired sources with small parallaxes, to avoid the need to correct for parallactic motion. 
Beginning with the full TGAS catalog of $\sim$2 million sources, we restricted to objects with PHOT\_G\_MEAN\_MAG $>$ 11.9. We also required the PARALLAX $<$ 1 mas and
PARALLAX\_ERROR $<$ 1 mas (i.e. parallax well-constrained to be very small). We further excluded low Galactic latitudes, demanding $|b_{gal}| > 15^{\circ}$. We also downselected
to sources with relatively small proper motions, resulting in a sample with median total proper motion of $|\mu| \approx 7$ mas/yr. We additionally required that the
proper motions in both coordinates be very well-measured, demanding PMRA\_ERROR $<$ 1 mas/yr and PMDEC\_ERROR $<$ 1 mas/yr. Not all of these cuts were strictly necessary. For instance, given the high quality of TGAS proper motion measurements, we likely could have made use of TGAS sources with large proper motions provided that their corresponding uncertainties are sufficiently small. Some of our cuts were made in the interest of paring down our TGAS sample to a similar size as that of the ICRF2 quasar sample -- we very likely could have obtained a much larger TGAS comparison sample had we desired one. Our cuts result in a sample of $\sim$1,900 TGAS sources.

The WISE counterparts to these TGAS sources have median magnitudes of 11.0 Vega in both W1 and W2, and have median (W1$-$W2) color of $-0.06$. The 25$^{th}$--75$^{th}$
percentile magnitude ranges are 10.1 $<$ W1 $<$  11.6 and 10.2 $<$ W2 $<$ 11.5. These ranges of magnitudes
are appropriate for evaluating the recalibrated astrometry, because these sources have very high signal-to-noise yet remain several magnitudes too faint to saturate in their
cores. To test our astrometry, we cross-matched our sample of TGAS sources with our full-sky Source Extractor catalogs described in $\S$\ref{sec:se}, using a 5$''$ match radius. We then computed RA and Dec residuals of the Source Extractor cross-matches relative to the Gaia positions, using both the pre-recalibration L1b-based astrometry and the SCAMP-based  astrometry which includes recalibration. In doing so, we accounted for proper motions of Gaia sources using the MJDs corresponding to the Source Extractor detections. RA residuals are in true angular units, i.e. we have multiplied the difference in RA by cos(Dec). The results are summarized in Figures \ref{fig:tgas_w1_raw}, \ref{fig:tgas_w1_recalib}, \ref{fig:tgas_w2_raw}, and \ref{fig:tgas_w2_recalib}. Each such plot is broken down into four subpanels. The top row shows pre-hibernation coadd epochs, while the bottom row shows post-hibernation epochs. The left column shows
residuals in RA and the right column shows residuals in Dec. The non-recalibrated residuals in Figures \ref{fig:tgas_w1_raw} and \ref{fig:tgas_w2_raw} display the bimodality explained in $\S$\ref{sec:astrom}, which results from converting our flux-weighted centroid measurements to world coordinates via the L1b WCS derived using asymmetric PSF models. On the
other hand, the recalibrated residuals shown in Figures \ref{fig:tgas_w1_recalib} and \ref{fig:tgas_w2_recalib} are very tightly distributed, always showing a single peak very nearly centered at zero. This indicates that our SCAMP recalibrations were quite successful in general. The robust standard deviation (1.4826$\times$ median absolute deviation) of the recalibrated residuals is 46 (45) mas in W1 and W2. We find this level of astrometric agreement versus an independent truth sample to be highly encouraging, as these values correspond to approximately 1/60 of a pixel and 1/140 of a FWHM.




\subsubsection{ICRF2 Quasars}
\label{sec:icrf}

We select a set of ICRF2 quasars to compare against our recalibrated astrometry, beginning with the full \cite{icrf2} catalog of 3,414 objects. We restrict
to the subset that have AllWISE counterparts within a 5$''$ radius. We further downselect to objects that have AllWISE \verb|cc_flags| = 0000, meaning they are not flagged as potentially
affected by WISE artifacts. This yields a comparison sample of $\sim$2,100 ICRF2 quasars. 

The median W1 (W2) magnitude of these sources is 14.6 (13.5), and the median color is (W1$-$W2) = 1.08. The ICRF2 counterpart magnitudes span 13.8 $<$ W1 $<$ 15.3, 12.7 $<$ W2 $<$ 14.2 from 25$^{th}$ to 75$^{th}$ percentile. There are thus two important distinctions between the ICRF2 and TGAS samples. First the ICRF2 sample is much fainter in both W1 and W2 than the TGAS sample. Therefore, we should expect the distributions of ICRF2 astrometric residuals to be much broader than those of the TGAS residuals, simply due to 
statistical scatter in the ICRF2 Source Extractor centroids. Second, whereas the TGAS sample is of typical stellar color, the ICRF2 sample is very red in (W1$-$W2), as expected
of quasars.

As for our TGAS sample, we cross-match our ICRF2 sample with our Source Extractor catalogs of $\S$\ref{sec:se} using a 5$''$ radius. We then compute the residuals in
each of RA and Dec, and again show plots broken down by WISE band and time period (pre versus post hibernation) in Figures \ref{fig:icrf_w1_raw}, \ref{fig:icrf_w1_recalib}, 
\ref{fig:icrf_w2_raw}, and \ref{fig:icrf_w2_recalib}. The most striking feature of these plots is that, even accounting for our SCAMP recalibrations, there remains some bimodal ``splitting''
of the residuals in forward-pointing versus backward-pointing scans. This splitting is more pronounced in W2 than in W1, and has a larger component along
the Dec direction than the RA direction in W2.

\subsubsection{Chromatic Astrometric Bias}
\label{sec:chromatic}
\begin{figure*}
       \includegraphics[width=7.0in]{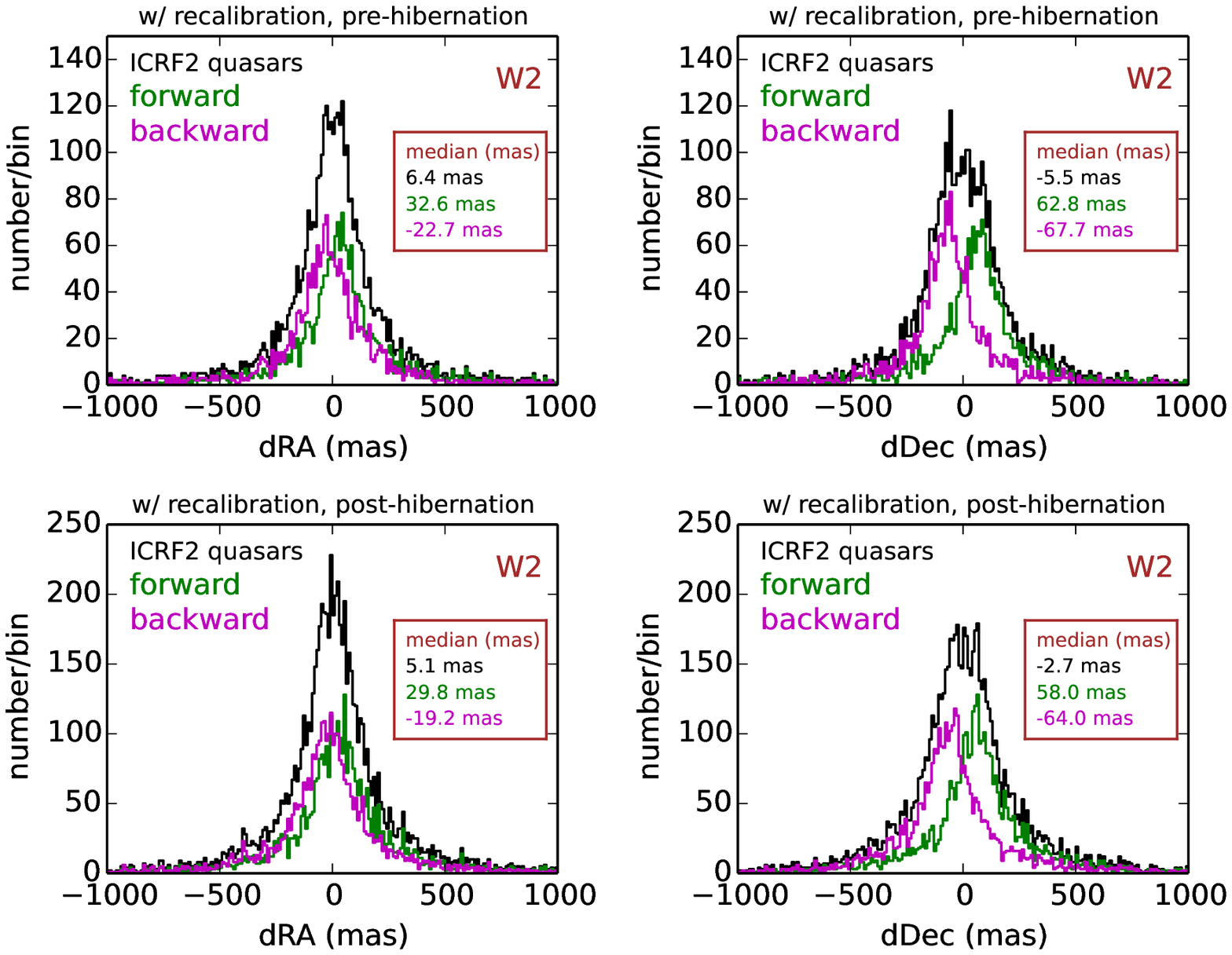}
   \caption{Same as Figure \ref{fig:icrf_w2_raw}, but showing residuals of our recalibrated ICRF2 source positions. The black distributions are sharper than in Figure \ref{fig:icrf_w2_raw}, but the forward-pointing/backward-pointing sub-distributions display some remaining separation, which we suggest may be due to a chromatic astrometric
   bias affecting these relatively red quasars ($\S$\ref{sec:chromatic}). }
    \label{fig:icrf_w2_recalib}
\end{figure*}

We hypothesize that the post-recalibration forward/backward splitting of the ICRF2 quasar residuals seen in Figures \ref{fig:icrf_w1_recalib} and \ref{fig:icrf_w2_recalib} is due to 
a chromatic astrometric bias. This solution is appealing because it can explain why the post-recalibration residuals of the TGAS sources are unimodal, whereas
those of the ICRF2 quasars are bimodal, with peaks corresponding to each scan direction: the TGAS sources have nearly the same typical color as our astrometric calibrators, 
whereas the IRCF2 quasars have a very different typical color. The splitting of residuals between forward and backward scans displayed in Figures \ref{fig:icrf_w1_recalib} and \ref{fig:icrf_w2_recalib} can be explained by a spectrum-dependent astrometric bias with a direction that is fixed \textit{relative to the detector}. The 180$^{\circ}$ rotation
of the detector between forward/backward pointing scans would then result in astrometric residuals like those seen, with two sub-distributions that appear to be
mirror images of each other about zero. To explain the ICRF2 residuals, the required amplitude of this effect would need be $\sim$68 mas in W2 and $\sim$28 mas in W1.







We can test this spectrum-dependent bias hypothesis by breaking the ICRF2 quasars into subsets according to color. Unfortunately, the distribution of W1$-$W2 colors among
the ICRF2 quasars is fairly narrow. The bluest 20\% of these quasars have median(W1$-$W2) = 0.77, whereas the reddest 20\% have median(W1$-$W2) = 1.29. We quantify the forward/backward splitting by computing the difference between the median of the forward residuals and the median of the backward residuals. The Dec splitting thus quantified
for the blue (red) subsample in W2 is 95 (145) mas. Thus, the redder subsample shows 52\% more splitting. This is qualitatively consistent with the hypothesis
that we are observing a chromatic bias which becomes stronger as the source spectrum becomes redder. Extrapolating the linear relation between W1$-$W2 color and W2 Dec splitting implied by the blue and red subsets, we would predict 17 mas of splitting at W1$-$W2 = $-0.04$, the median color of our astrometric calibrators. This is reasonably
consistent with zero splitting at the calibrator color (as seen with the TGAS sample in Figures \ref{fig:tgas_w1_recalib} and \ref{fig:tgas_w2_recalib}), especially
considering the small number of unique quasars in the red/blue subsets and the fact that there is no clear physical basis for extrapolating linearly in astrometric splitting per
magnitude of W1$-$W2 color.

Another appealing aspect of the spectrum-dependent astrometric bias hypothesis is that such a bias could be time-independent and still explain the
post-recalibration astrometric residuals. This can be seen by observing that each pair of subplots directly above/below one another
in Figures \ref{fig:icrf_w1_recalib} and \ref{fig:icrf_w2_recalib} shows a consistent amplitude of the astrometric splitting between forward/backward scans.

Because the TGAS and ICRF2 samples have very different brightnesses in addition to different typical colors, one might ask whether the astrometric bias could
be brightness-dependent rather than spectrum-dependent. However, the hypothesis of a brightness-dependent astrometric bias is qualitatively inconsistent
with our findings. The ICRF2 quasars are much more similar in brightness to our astrometric calibrator sample than are the TGAS sources (the TGAS sources
are several magnitudes brighter than our typical calibrators). Thus, in the brightness-dependent bias scenario, the TGAS residuals after recalibration would be more
bimodal than the ICRF2 residuals after recalibration, which is the opposite of the behavior we observe.

\begin{figure}
        \includegraphics[width=3.3in]{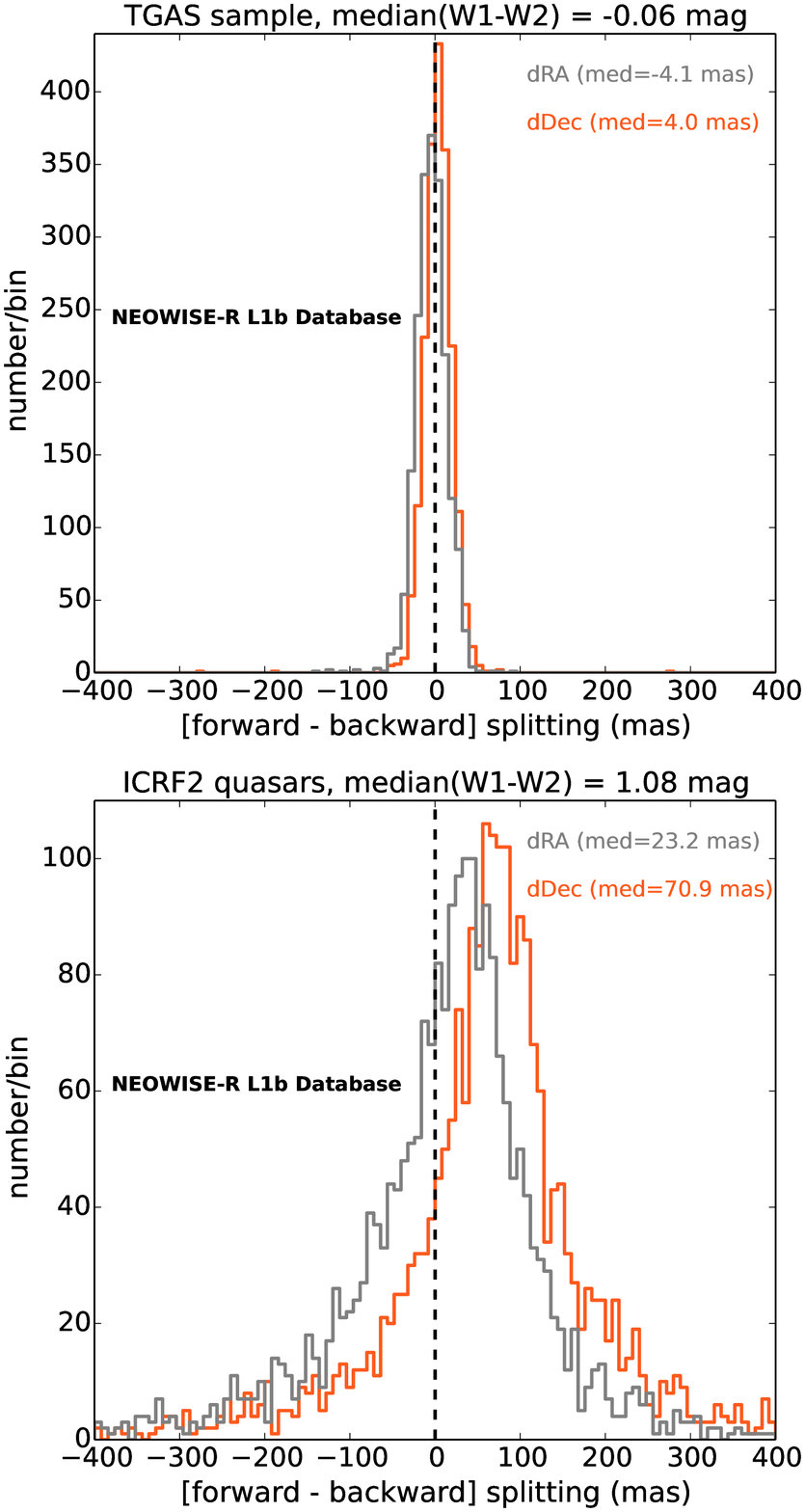}
    \caption{Histograms of per-object, per-coordinate astrometric splittings between forward-pointing and backward-pointing scans as derived from astrometry
    in the NEOWISE-R Single Exposure Source Table. Top: Our TGAS sample of $\S$\ref{sec:tgas}. The distributions of splittings in RA (grey) and Dec (orange) are both 
    centered very nearly at zero, indicating no systematic biases between scan directions for these normal-colored sources. Bottom: Splittings for our ICRF2 quasar sample of $\S$\ref{sec:icrf}, which are very red relative to typical astrometric calibrator sources. The distributions are off-center, consistent with our hypothesis of a chromatic astrometric bias
    with a direction that is fixed relative to the detector. The ICRF2 distributions are relatively broad because these objects are much fainter than those of our TGAS sample.}
    \label{fig:l1b_shifts}
\end{figure}

We have also analyzed  astrometry drawn from the NEOWISE-R Single Exposure Source Table for both our TGAS and ICRF2 samples, to determine whether the 
hypothesized chromatic
bias is present there as well. We queried IRSA for all NEOWISE-R Single Exposure Source Table astrometry for every source in our TGAS and ICRF2 samples. 
Typically, for sources brighter than the single-exposure detection limit, there are $\sim$70 single-frame epochs of astrometry available. For each source, we
compute its median L1b source table position in forward-pointing scans and in backward-pointing scans. For each source, we then compute its 
astrometric splitting as the difference between these two median positions (each source's astrometric splitting can be broken down into components in RA and Dec). The results are plotted in Figure \ref{fig:l1b_shifts}. 
The top panel shows histograms of the per-object astrometric splittings for our TGAS sample. Because these distributions are centered very close to zero, there appears to be no systematic positional splitting between 
forward/backward scans, down to the level of a handful of mas. This is consistent with our results shown in Figures \ref{fig:tgas_w1_recalib} and \ref{fig:tgas_w2_recalib}, which
indicate sharp unimodal astrometric residuals for our TGAS sample. The bottom panel of Figure \ref{fig:l1b_shifts} shows histograms of the per-object
astrometric splittings for our ICRF2 sample. The astrometric splitting distributions are clearly shifted 
relative to zero, indicating systematic offsets between the WISE team's measurements of ICRF2 quasar positions in forward versus backward scans. The 
sign of the splitting is consistent with that seen in Figures \ref{fig:icrf_w1_recalib} and \ref{fig:icrf_w2_recalib}. Also, just as we saw with our unWISE-based measurements,
the W2 L1b-based astrometric splitting measurement indicates a larger amplitude in the Dec direction than in the RA direction. It is difficult to perform an exact quantitative comparison between the 
ICRF2 astrometric splittings measured from our unWISE coadds and those measured from the WISE team's L1b database, because the unWISE centroids are calculated independently in each band
while the L1b database centroids are calculated jointly in W1 and W2. In summary, we believe that the behavior displayed by the NEOWISE-R Single Exposure Source Table astrometry analyzed is consistent
with our hypothesis of a chromatic astrometric bias.

Admittedly, we do not have a detailed physical explanation for the origin of this hypothesized chromatic bias in terms of the WISE instrumentation. 
We note that, so long as the bias is analyzed in terms of forward versus backward astrometric splittings of individual objects, one need not make any reference to high-fidelity ground truth coordinates 
like those available for our TGAS and ICRF2 samples. Therefore, it should be possible to test/study this effect with a much larger sample of quasars, for example all mid-IR bright quasars
spectroscopically confirmed by SDSS/BOSS. With this dramatically expanded sample, it may be possible to more fully map out the chromatic bias as a function of
source spectrum and/or make a phenomenological model suitable for computing astrometric corrections as a function of source spectrum.

Lastly, we note that the possibility of a chromatic astrometry bias affecting sources with red W1$-$W2 colors raises concerns about the prospect using our 
time-resolved coadds (or any WISE data) to measure parallaxes for cold brown dwarfs detected primarily at W2. Because the scan direction at a given sky location flips when WISE observes it from opposite sides
of the Sun, there is a sense in which the astrometric bias behaves as a spurious `parallax'. While the $\lesssim 70$ mas amplitude of the proposed bias is very small
relative to e.g. the WISE FWHM ($\lesssim 0.01$ FWHM), it is non-negligible relative to the parallaxes of all but the most nearby sources. Complicating matters,
any chromatic astrometric bias must be a function of the \textit{in-band} source spectrum within each WISE channel rather than simply the W1$-$W2 color.









\subsection{Photometric Repeatability}
We have also sought to characterize the photometric repeatability of our time-resolved coadds. By photometric repeatability, we mean the standard deviation of a
given source's flux values measured independently in multiple epochal coadds, expressed as a fraction of that source's average flux across all epochs.

We use the AllWISE catalog to select samples of sources with which to evaluate the repeatability in each band. The ideal objects for assessing photometric repeatability will be 
bright point sources which remain unsaturated in their cores. The approximate saturation thresholds are\footnote{\scriptsize{\url{http://wise2.ipac.caltech.edu/docs/release/allwise/expsup/sec2_1.html}}} W1 $\approx$ 8.0 and W2 $\approx$ 7.0. We therefore select samples from the AllWISE catalog with 9.5 $\le$ \verb|w1mpro| $\le$ 10.5 for W1 and 8.7 $\le$ 
\verb|w2mpro| $\le$ 9.7 for W2. We also demand a 2MASS counterpart within 1$''$, so that we can neglect proper motion when photometering our coadd epochs which span over half a decade. We further restrict to high Galactic latitude ($|b_{gal}| > 60^{\circ}$) and low ecliptic latitude ($|\beta| < 30^{\circ}$) to avoid confusion/crowding. Finally, we require the AllWISE parameter \verb|nb| to be 1 (source is not blended) and \verb|w?cc_map| to be 0 in the relevant band (no artifact flags). Applying these criteria, we obtain samples of $\sim$35,000 ($\sim$18,000) unique sources in W1 (W2).

For each such source, we identify the \verb|coadd_id| footprint in which it appears furthest from the tile boundary. We then measure its fluxes in all corresponding epochal coadds for 
which the source's location has a minimum integer coverage of 12 frames. In detail, the photometry is performed with the \verb|djs_phot| aperture photometry routine\footnote{\scriptsize{\url{http://www.sdss3.org/dr8/software/idlutils_doc.php\#DJS_PHOT}}}, using
a 19.25$''$ aperture radius, an inner sky radius of 55$''$ and an outer sky radius of 74.25$''$. We opt to use aperture photometry rather than PSF photometry because 
the W1/W2 PSFs have broadened by a few percent over the course of WISE's lifetime \citep{cutri15}, and we
have not yet constructed time-dependent PSF models appropriate for the unWISE epochal coadds. We further restrict our sample to those sources which have
at least five coadd epochs worth of aperture photometry measured in the relevant WISE band. For each unique source in each band, we compute the standard deviation and mean of our measured
fluxes, with the repeatability taken to be the ratio of these two quantities. Averaging over all sources, we find mean repeatabilities of 1.8\% (1.9\%) in W1 (W2). Our repeatability estimates does not attempt to account for any contribution from intrinsic variability of the sources we photometered, as we assume this to be a negligible factor.

This level of repeatability is certainly sufficient for identifying interesting cases of extreme variability (e.g. the object shown in Figure \ref{fig:quasar}). However, because
WISE is a space-based mission delivering very uniform image quality, we find 1.8--1.9\% repeatability to be disappointing. These values are far too large to be dominated by statistical
noise or imperfections in background level estimation, given the bright nature of the sources analyzed. Restricting to the fainter half of sources photometered in each band
leads to negligible changes in our repeatability estimates, suggesting that these have not been inflated by pushing our samples too aggressively toward the onset of saturation.  In the future we will attempt to further investigate and/or improve the photometric repeatability.




\subsection{Full-depth Photometry Comparison}
We have additionally checked that photometry of our time-resolved coadds is consistent with photometry of our full-depth unWISE coadds. The only publicly available photometry
of the time-resolved coadds presented in this work is forced photometry from Data Release 4 (DR4) of the DESI imaging Legacy Survey \citep{dey2018}.
This DR4 forced photometry covers $\sim$3,600 square degrees of extragalactic sky in the north Galactic cap ($\delta \gtrsim 30^{\circ}$). To compare
full-depth versus time-resolved forced photometry, we select all sources from DR4 satisfying the following criteria:

\begin{itemize}
\item Full-depth forced photometry signal-to-noise between 50 and 1000 in the relevant WISE band.
\item No unWISE bright star mask bits set (\verb|WISEMASK_W?| = 0) in the relevant WISE band.
\item No flags set in the optical (\verb|ANYMASK_G| = 0, \verb|ANYMASK_R| = 0, \verb|ANYMASK_Z| = 0).
\item Isolated (\verb|FRACFLUX_W?| $<$ 0.01) in the relevant WISE band.
\end{itemize}

These cuts are applied independently for each WISE band, yielding a sample of 2.54 million (0.81 million) sources in W1 (W2). From 1$^{st}$ to 99$^{th}$ percentile, these samples 
span 13.0 $<$ W1 $<$ 16.4 and 11.8 $<$ W2 $<$ 15.1. For each source in its relevant WISE band, we compute the inverse variance weighted mean of the available flux measurements based on our time-resolved coadds, which we refer to as the `lightcurve mean flux'. Typically, forced photometry measurements from six coadd epochs contribute to the lightcurve mean flux calculation for each source in each band. The median ratio of the full-depth flux to the lightcurve mean flux is 1.008 (1.007) with a standard deviation of 0.009 (0.006) in W1 (W2). We have not investigated the origin of the apparent $\lesssim$1\% bias whereby full-depth fluxes seem to be systematically larger
than lightcurve mean fluxes. It is possible that this bias results from our inverse variance weighting when averaging lightcurve fluxes, as Poisson noise makes variance correlated with flux. Nevertheless, the fluxes measured from epochal coadds appear to trace those measured in the deeper stacks quite well. \cite{meisner17} established that the full-depth forced photometry from DR4 is consistent with the AllWISE catalog at the sub-percent level in terms of multiplicative calibration, indicating that our time-resolved coadd photometry is also consistent on average with AllWISE at roughly the percent level over the magnitude range analyzed.



\section{Data Release}
\label{sec:dr}




\subsection{Detailed Content/Format}
Our time-resolved coadds are available for download online\footnote{\url{http://unwise.me/tr_neo2}}. The top level of the data release hierarchy contains: 

\begin{enumerate}
\item A set of 112 directories named \verb|e000|, \verb|e001|, ... \verb|e111|. Each such directory contains all of the coadd outputs corresponding to a single \verb|epoch| value, with
the \verb|epoch| value encoded as a zero-padded three-digit integer in the final three characters of the \verb|e???| directory name. Within each \verb|e???| directory, 
the coadd hierarchy is the same as the full-depth coadd hierarchy employed in previous unWISE data releases. For example, all outputs for (\verb|coadd_id|, \verb|epoch|) = (3524m031, 4) are in the directory \verb|e004/352/3524m031|, including both W1 and W2. Every FITS image output includes header keywords MJDMIN and MJDMAX indicating the range of MJDs spanned by the contributing exposures. The entire set of coadd image outputs is 15.5 TB in size.
\item A file called \verb|tr_neo2_index.fits|, which can be used to determine which coadd epochs are available for each \verb|coadd_id| in each band. This file
contains a FITS binary table, wherein each row provides metadata about an existing set of unWISE coadd outputs for a single (\verb|coadd_id|, \verb|epoch|, \verb|band|) triplet.
Table \ref{tab:index} lists the columns present in this index file and their meanings.
\item A file named \verb|tr_neo2_scamp_headers.tar.gz| that contains all of the SCAMP headers derived during the astrometric recalibration analysis presented in $\S$\ref{sec:astrom}. Users interested in 
performing astrometry with flux-weighted centroids from e.g. Source Extractor may find these helpful. Note that the final nine columns of each row of the \verb|tr_neo2_index.fits| metadata table contain
all information necessary to reconstruct the corresponding SCAMP WCS solution, without needing to access the .head header files individually.
\end{enumerate}

Some users may not wish to download large volumes of image data. We note that the need to download our data release files can be circumvented with a browser-based visualization tool called WiseView\footnote{\url{http://byw.tools/wiseview}}, which renders image blinks of our time-resolved coadds according to user specified parameters including sky location, colormap and stretch, and frame rate.

\begin{table*}
        \centering
        \caption{Epochal coadd index table column descriptions.}
        \label{tab:index}
        \begin{tabular}{ll}
                \hline
                Column & Description \\
                 \hline
                COADD\_ID & \verb|coadd_id| astrometric footprint identifier as defined in $\S$\ref{sec:tiling} \\
                BAND & integer WISE band; either 1 or 2  \\
                EPOCH & \verb|epoch| number, as defined in $\S$\ref{sec:slicing} \\
                RA & tile center right ascension (degrees) \\
                DEC & tile center declination (degrees) \\
                FORWARD & boolean indicating whether input frames were acquired pointing forward or backward along Earth's orbit \\
                MJDMIN & MJD value of earliest contributing exposure \\
                MJDMAX & MJD value of latest contributing exposure \\
                MJDMEAN & mean of MJDMIN and MJDMAX \\
                DT & difference of MJDMAX and MJDMIN (days) \\
                N\_EXP & number of exposures contributing to the coadd \\
                COVMIN & minimum integer coverage in unWISE \verb|-n-u| coverage map \\
                COVMAX & maximum integer coverage in unWISE \verb|-n-u| coverage map \\
                COVMED & median integer coverage in unWISE \verb|-n-u| coverage map \\
                NPIX\_COV0 & number of pixels in \verb|-n-u| map with integer coverage of 0 frames \\
                NPIX\_COV1 & number of pixels in \verb|-n-u| map with integer coverage of 1 frame \\
                NPIX\_COV2 & number of pixels in \verb|-n-u| map with integer coverage of 2 frames \\
                LGAL & Galactic longitude corresponding to tile center (degrees) \\
                BGAL & Galactic latitude corresponding to tile center (degrees) \\
                LAMBDA & ecliptic longitude corresponding to tile center (degrees) \\
                BETA & ecliptic latitude corresponding to tile center (degrees) \\
                SCAMP\_HEADER\_EXISTS & boolean indicating whether SCAMP astrometric recalibration succeeded \\
                SCAMP\_CONTRAST & SCAMP contrast parameter; $-1$ when SCAMP\_HEADER\_EXISTS is 0 \\
                ASTRRMS1 & SCAMP ASTRRMS1 output parameter; units of degrees; 0 when SCAMP\_HEADER\_EXISTS is 0 \\
                ASTRRMS2 & SCAMP ASTRRMS2 output parameter; units of degrees; 0 when SCAMP\_HEADER\_EXISTS is 0 \\
                N\_CALIB & number of sources used for SCAMP astrometric recalibration \\
                N\_BRIGHT & number of `bright' (S/N $>$ 100) sources among astrometric calibrators \\
                N\_SE & total number of Source Extractor objects in this coadd \\
                NAXIS & 2-elements NAXIS array from SCAMP solution; 0 when SCAMP\_HEADER\_EXISTS is 0 \\
                CD & 2$\times$2 CD matrix from SCAMP solution; NaNs when SCAMP\_HEADER\_EXISTS is 0 \\
                CDELT & 2-element CDELT array from SCAMP solution; NaNs when SCAMP\_HEADER\_EXISTS is 0 \\
                CRPIX & 2-element CRPIX array from SCAMP solution; NaNs when SCAMP\_HEADER\_EXISTS is 0 \\
                CRVAL & 2-element CRVAL array from SCAMP solution; NaNs when SCAMP\_HEADER\_EXISTS is 0 \\
                CTYPE & 2-element CTYPE array from SCAMP solution; empty strings when SCAMP\_HEADER\_EXISTS is 0 \\
                LONGPOLE & LONGPOLE parameter from SCAMP solution; NaNs when SCAMP\_HEADER\_EXISTS is 0 \\
                LATPOLE & LATPOLE parameter from SCAMP solution; NaNs when SCAMP\_HEADER\_EXISTS is 0 \\
                PV2 & 2-element PV2 array from SCAMP solution; NaNs when SCAMP\_HEADER\_EXISTS is 0 \\
                \hline
        \end{tabular}
\end{table*}



\subsection{Cautionary Notes}

Here we address a list of of potential stumbling blocks which users may encounter and/or find counterintuitive. All of these issues are considered to be features rather
than bugs. In the event that we become aware of any bugs we will attempt to document them on the data release website.


\begin{figure*}
        \includegraphics[width=7.0in]{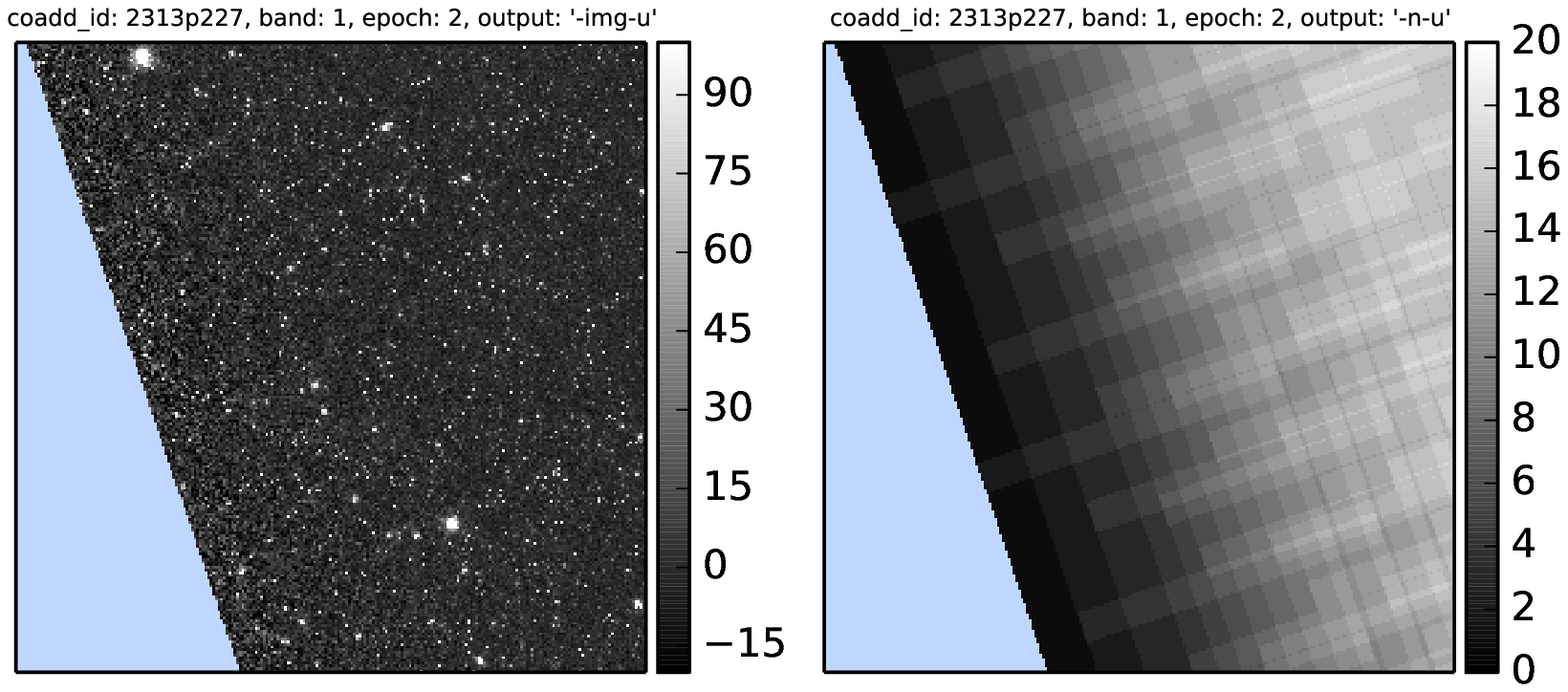}
    \caption{Example of a so-called `partial coadd'. Left: rendering of unWISE \texttt{-img-u} output. Colorbar units are Vega nanomaggies. Right: rendering of corresponding unWISE
    \texttt{-n-u} integer coverage map. Colorbar units are number of exposures. The light blue indicates pixels with zero integer coverage i.e. missing data. This coadd has a region with missing data because WISE went into hibernation partway through this visit.}
    \label{fig:partial}
\end{figure*}

\begin{enumerate}
\item Many coadd images have extended regions with zero integer coverage i.e. missing data or `holes'. We refer to such cases as `partial coadds'. There are several
situations that can lead to partial coadds. For example, coadd epochs that begin just before the pre-hibernation NEOWISE mission ends are often cut short by the
associated halting of survey operations and therefore have large empty regions. Figure \ref{fig:partial} shows one such example. 2.2\% (98.3\%) of epochal coadds at $|\beta| < 80^{\circ}$  ($|\beta| >80^{\circ}$) have missing regions. Our segmentation of frames into $<$ 15 day long epochs near the ecliptic poles (described in $\S$\ref{sec:slicing}) means that each such epoch samples a relatively small range of approach angles toward the pole. As a result, the regions of nonzero coverage in these coadds are stripes narrower than the 1.56$^{\circ}$ unWISE tile angular size. One could imagine adopting a policy wherein we only keep coadd outputs for which all pixels have at least some minimum integer coverage, with the threshold value being larger than zero frames. We believe that such an approach would unnecessarily discard useful data, for example the entire western half of the partial coadd shown in Figure \ref{fig:partial}. Empty portions of coadds can be identified by checking for regions within the unWISE \verb|-n-u| integer coverage maps that have values of 0.
\item Users should also exercise caution in regions with very low but nonzero integer coverage. For instance, regions of epochal coadds with integer coverage $\le 2$ do not 
benefit from the usual unWISE per-pixel outlier rejection. This outlier rejection works by inter-comparing the resampled L1b pixel values contributing to a given pixel in coadd space. But with only $\le 2$
resampled L1b values it is not possible to determine which if any of the L1b images contribute an outlier. 
\item As previously mentioned in $\S$\ref{sec:slicing}, it is possible for a (\verb|coadd_id|, \verb|band|) pair to have coadd outputs for \verb|epoch| = $(N+1)$ but not \verb|epoch| = $N$, with $N \ge 0$. For instance, (\verb|coadd_id| = 0016m213, \verb|band| = 2) has coadd outputs for \verb|epoch| = 0, 1, 3, 4, 5, 6 but not \verb|epoch| = 2.
\item There are rare cases in which a (\verb|coadd_id|, \verb|epoch|) pair has coadd outputs in only one of the two WISE bands. One such case is (\verb|coadd_id| = 1929p045, \verb|epoch| = 0), which has coadd outputs for W1 but not W2.
\item As previously discussed in $\S$\ref{sec:slicing}, it is not safe to assume that \verb|epoch| = $N$ of one \verb|coadd_id| has a mean MJD similar to \verb|epoch| = $N$ of
a different \verb|coadd_id|.
\item Although the mean MJDs in W1 and W2 are usually quite similar for a given (\verb|coadd_id|, \verb|epoch|) pair, a small number of counterexamples exist. For instance, the (\verb|coadd_id|, \verb|epoch|) = (2708p636, 37) coadd outputs have a $\sim$120 day offset between the mean MJDs in W1 and W2.
\item The bottom-level coadd output file names for a given (\verb|coadd_id|, \verb|band|) pair are the same for all epochs. For example, the W1 \verb|-img-u| unWISE output images
for \verb|coadd_id| = 0000p000 are named \verb|unwise-0000p000-w1-img-u.fits| for both \verb|epoch| = 0 and \verb|epoch| = 1. These two files reside in directories called \verb|e000/000/0000p000| and \verb|e001/000/0000p000|, respectively, so that their file names are 
unique when including the full paths within our data release hierarchy.
\item The recalibrated WCS solutions described in $\S$\ref{sec:scamp} are \textit{not} contained within the FITS headers of our time-resolved coadd image files. The recalibrated
WCS solutions can be obtained from either \verb|tr_neo2_scamp_headers.tar.gz| or \verb|tr_neo2_index.fits|, as described in $\S$\ref{sec:dr}.
\item Caution should be exercised when using our SCAMP WCS solutions in sky regions with very high source density, e.g. the Galactic center and the LMC. The 
SCAMP solutions rely on our Source Extractor catalogs described in $\S$\ref{sec:se}, and these are likely to be severely corrupted by deblending failures in crowded fields. Large values of the ASTRRMS1, ASTRRMS2 parameters in our \verb|tr_neo2_index.fits| metadata table can serve as indicators of questionable SCAMP solutions, as can unusually low values of N\_CALIB.
\item SCAMP solutions failed for 639 time-resolved coadds (0.27\% of all coadds). These failures are all associated with partial coadds that have exceptionally large missing
regions and therefore yield few Source Extractor detections suitable for use by SCAMP. Such cases are flagged with
\verb|SCAMP_HEADER_EXISTS| = 0 in the \verb|tr_neo2_index.fits| metadata table.
\end{enumerate}





\section{Conclusion}
\label{sec:conclusion}

We have created the first ever full-sky set of time-resolved W1/W2 coadds. At typical sky locations, six such epochal coadds spanning a 5.5 year time baseline have been generated in each band.
These coadds will enable major breakthroughs in studying motions and long-timescale variability of WISE sources far below the single-exposure detection limit.

Much work still remains in order to fully exploit the combined WISE+NEOWISE imaging data set. In particular, by incorporating all NEOWISER frames through the end of 2017, it will be possible to increase the typical number of unWISE coadd epochs per sky location from six to ten, extending the time baseline from 5.5 to 7.5 years. It will also be critical to translate our
unWISE coadd images into a full-sky, WISE-selected catalog. The CatWISE program (PI: Eisenhardt) is combining the full-depth and time-resolved unWISE coadds to produce motion-aware
W1/W2 catalogs similar to AllWISE, but leveraging the greatly increased time baseline and much deeper source extractions enabled by post-reactivation imaging. One could also imagine generating light curves for all AllWISE (or CatWISE) sources by performing forced photometry
on our set of time-resolved unWISE coadds, thereby obtaining a deep/clean all-sky database of long-timescale variability in the mid-IR.


\section*{Acknowledgements}

This work has been supported by grant NNH17AE75I from the NASA Astrophysics Data Analysis Program. We thank Dan Caselden and Paul Westin for implementing, hosting, and maintaining the WiseView visualization tool.

This research makes use of data products from the Wide-field Infrared Survey Explorer, which is a joint project of the University of California, Los Angeles, and the Jet Propulsion Laboratory/California Institute of Technology, funded by the National Aeronautics and Space Administration. This research also makes use of data products from NEOWISE, which is a project of the Jet Propulsion Laboratory/California Institute of Technology, funded by the Planetary Science Division of the National Aeronautics and Space Administration. This research has made use of the NASA/ IPAC Infrared Science Archive, which is operated by the Jet Propulsion Laboratory, California Institute of Technology, under contract with the National Aeronautics and Space Administration.

The National Energy Research Scientific Computing Center, which is supported by the Office of Science of the U.S. Department of Energy under Contract No. DE-AC02-05CH11231, provided staff, computational resources, and data storage for this project.

\bibliographystyle{mnras}
\bibliography{tr_neo2}

\end{document}